\documentstyle[epsf,12pt,aaspp4]{article}

\def\etal{et~al.}

\def\ms{\hbox{ms}}
\def\cm{\hbox{cm}}
\def\km{\hbox{km}}
\def\s{\hbox{s}}
\def\ergs{\hbox{ergs}}
\def\keV{\hbox{keV}}
\def\MeV{\hbox{MeV}}
\def\Mpc{\hbox{Mpc}}

\lefthead{Brainerd}
\righthead{Cosmological Signatures of Gamma-Ray Bursts}

\begin{document}

\title{The Cosmological Signatures and Flux Distribution\\
of Gamma-Ray Bursts with a Broad Luminosity Distribution\altaffilmark{1}}

\author{J. J. Brainerd\altaffilmark{2}}
\affil{University of Alabama in Huntsville}

\altaffiltext{1}{To appear in The Astrophysical Journal, Sept. 20, 1997}
\altaffiltext{2}{e-mail: jim.brainerd@msfc.nasa.gov}

\authoraddr{Space Sciences Lab, ES-84, NASA/Marshall Space Flight Center,
Huntsville, AL \ 35812\\}
\authoremail{jim.brainerd@msfc.nasa.gov\\}

\cpright{AAS}{1997}
\slugcomment{To appear in the Astrophysical Journal, Sept. 20, 1997}

\begin{abstract}

The cosmological expansion cannot produce the reported correlations
of the gamma-ray burst timescale and spectral energy with peak flux
if the burst model reproduces the BATSE 3B peak-flux distribution
for a non-evolving burst source density.
The required ratios of time dilation and redshift factors are only
produced by monoluminous models at peak fluxes below the BATSE threshold,
and they are never produced by power-law luminosity models.  Monoluminous
models only produce acceptable fits to the peak-flux distribution for
very specific combinations of the spectral and cosmological parameters.
The redshift of gamma-ray bursts at the BATSE threshold is $z \approx 1.5$.
Power law luminosity distribution models $\propto L^{-\beta}$ produce
acceptables fits to the data for most values of the spectral parameters
when $\beta < 1.6$.  In this model, gamma-ray bursts of a given peak flux
have a distribution of redshifts, with a maximum redshift of $\gtrsim 3$
for peak fluxes near the BATSE threshold, and with an average redshift
of $< 1$ for all values of peak flux.
This qualitative behavior occurs whenever the luminosity distribution
determines the shape of the peak-flux distribution, regardless of
whether source density evolution occurs.
The reported correlations of the burst timescale and the spectral energy
with peak flux are systematically 1 standard deviation above the
monoluminous model and 1.5 to 2 standard deviation above the power-law
luminosity model.  These results suggest that an intrinsic correlation
of  burst timescale and spectral energy with luminosity is present.
Studies of the peak-flux distribution for bursts selected by $E_{peak}$
or hardness ratio provide a test for this intrinsic correlation.

\end{abstract}


\newpage
\section{The Indistinctness of Cosmological Signatures}

Gamma-ray bursts are widely believed to be cosmological, but no direct
measurement of distance exists to confirm this belief.  No gamma-ray
burst has lines that can be associated with atomic or nuclear processes,
so a direct calculation of their redshifts is not possible.  With only
one recent exception (\markcite{Groot}Groot \etal\ 1997), no gamma-ray
burst has an observed counterpart, so an indirect calculation of their
redshifts is also not possible.  Because of these difficulties, a number
of researchers have attempted to derive average redshifts by studying
the properties of gamma-ray burst catalogs. The simplest of these studies
fit the peak-flux distribution to cosmological models under various
assumptions
(e.g. \markcite{Fenimore1} Fenimore \etal\ 1993;
\markcite{Wickramasinghe} Wickramasinghe \etal\ 1993;
\markcite{Emslie} Emslie \& Horack 1994;
\markcite{Cohen} Cohen \& Piran 1995;
\markcite{Pendleton1} Pendleton \etal\ 1996;
\markcite{Petrosian} Petrosian \& Lee 1997).
More involved studies attempt to measure a correlation of gamma-ray burst
timescale with photon flux under the assumption that such a correlation
is evidence of a cosmological time dilation
(\markcite{Norris1}Norris \etal\ 1994;
\markcite{Davis}Davis \etal\ 1994;
\markcite{Mitrofanov2}Mitrofanov \etal\ 1994,\markcite{Mitrofanov3} 1996
\markcite{Lee} Lee \& Petrosian 1996).
Related studies attempt to measure a correlation of the characteristic
photon energy with flux, which may indicate a cosmological redshift
(\markcite{Nemiroff}Nemiroff \etal\ 1994;
\markcite{Mallozzi1} Mallozzi \etal\ 1995).

The difficulty with these approaches is that the interpretation of the
results is not unique.  The observed flux distribution simply shows that
gamma-ray bursts are not homogeneously distributed in space,
which is unique neither to a cosmological origin nor to a particular
model of cosmological expansion.  Galactic corona models of gamma-ray
bursts easily reproduce the observations
(\markcite{Brainerd1}Brainerd 1992;
\markcite{Ulmer1}Ulmer \& Wijers 1995;
\markcite{Ulmer2}Ulmer \etal\ 1995;
\markcite{Hakkila1}Hakkila \etal\ 1995),
and any cosmological model can be made to conform to the
observations with the proper choice of density evolution
(\markcite{Rutledge}Rutledge \etal\ 1995;
\markcite{Horack1}Horack \etal\ 1995;
\markcite{Reichart}Reichart \& M\'esz\'aros 1997),
or luminosity distribution
(\markcite{Meszaros}M\'esz\'aros \& M\'esz\'aros 1995;
\markcite{Hakkila2} Hakkila \etal\ 1996),
as is evident from the numerous combinations of assumptions made
by the numerous papers examining this topic.  The effects sought in
correlation studies can be explained by an intrinsic correlation
of the parameter of interest with the luminosity
(\markcite{Brainerd2}Brainerd 1994a).
The beaming of relativistic jets
(\markcite{Brainerd2}Brainerd 1994a; \markcite{Yi} Yi 1994)
and the conservation of total energy in monoenergetic sources
(\markcite{Wijers}Wijers \& Paczy\'nski 1994) are examples of
mechanisms that produce such a correlation.

The question addressed in this article is whether a strong correlation of
the burst time scale or the characteristic spectral energy with peak flux
can have a purely cosmological origin when the gamma-ray burst peak-flux
distribution is determined by the luminosity distribution rather than by
the spatial distribution.  In this study, I use a luminosity function that
is a power law with an upper limit, and, for comparison, a monoluminous
distribution.  The spectral model I use (\S 2) is the Compton attenuation
model (\markcite{Brainerd3}Brainerd 1994b), which accurately represents
the observed spectra.  The cosmological expansion with a
$\left( 1 + z \right)^{\mu}$ comoving source density ($\mu < 9/2$)
defines a limiting power-law index for the peak flux distribution that
is $> -5/2$ as the peak flux goes to zero; a luminosity distribution
can drive this index closer to $-5/2$ (\S 3 and \S 4).  When the luminosity
function determines the asymptotic behavior of the peak flux distribution,
the density evolution of the gamma-ray burst sources has only a weak effect
on the average time-dilation and redshift (\S 3).  I find (\S 5) that
a broad luminosity distribution in a Friedmann cosmology with a constant
comoving source density produces peak-flux distributions that fit well
the distribution observed by the Burst and Transient Source Experiment
(BATSE) on the Compton Gamma-Ray Observatory.  At a given flux, the
gamma-ray bursts have a redshift distribution, with the average redshift
considerably smaller than the maximum redshift (\S 4).  The maximum redshift
a burst can have at the BATSE threshold is $\gtrsim 3$.  In this model,
the cosmological expansion cannot explain the reported correlations (\S 6),
so an intrinsic correlation must exist.  The flux distribution
for gamma-ray bursts selected by hardness ratio presented in
recent studies is a measure of this intrinsic correlation (\S 7).
These studies offer the opportunity to separate the cosmological
effects from the intrinsic correlations.


\section{Spectral Parameterization}

Gamma-ray burst spectra are not power-law spectra; they are
hard spectra that gradually roll over in the $50 \, \keV$
to $300 \, \keV$ energy band to a power law, with the power-law
behavior appearing above $1 \, \MeV$
(\markcite{Metzger}Metzger \etal\ 1974;
\markcite{Hueter}Hueter 1984;
\markcite{Matz}Matz \etal 1985).
The choice of spectral model influences the shape of the
burst peak-flux distribution (\markcite{Brainerd3}Brainerd 1994b;
\markcite{Mallozzi2}Mallozzi \etal\ 1996).
A physical spectral model that accurately describes gamma-ray burst
spectra is a Compton-attenuated power-law spectrum
(\markcite{Brainerd3}Brainerd 1994b;
\markcite{Brainerd4}\markcite{Brainerd5}Brainerd \etal\ 1996a,b).
Because this spectrum is a testable gamma-ray burst theory, it is of
interest beyond its descriptive value.

\begin{figure}[t]
\hbox to 6.5in {
\epsfxsize=4.0in \epsfbox{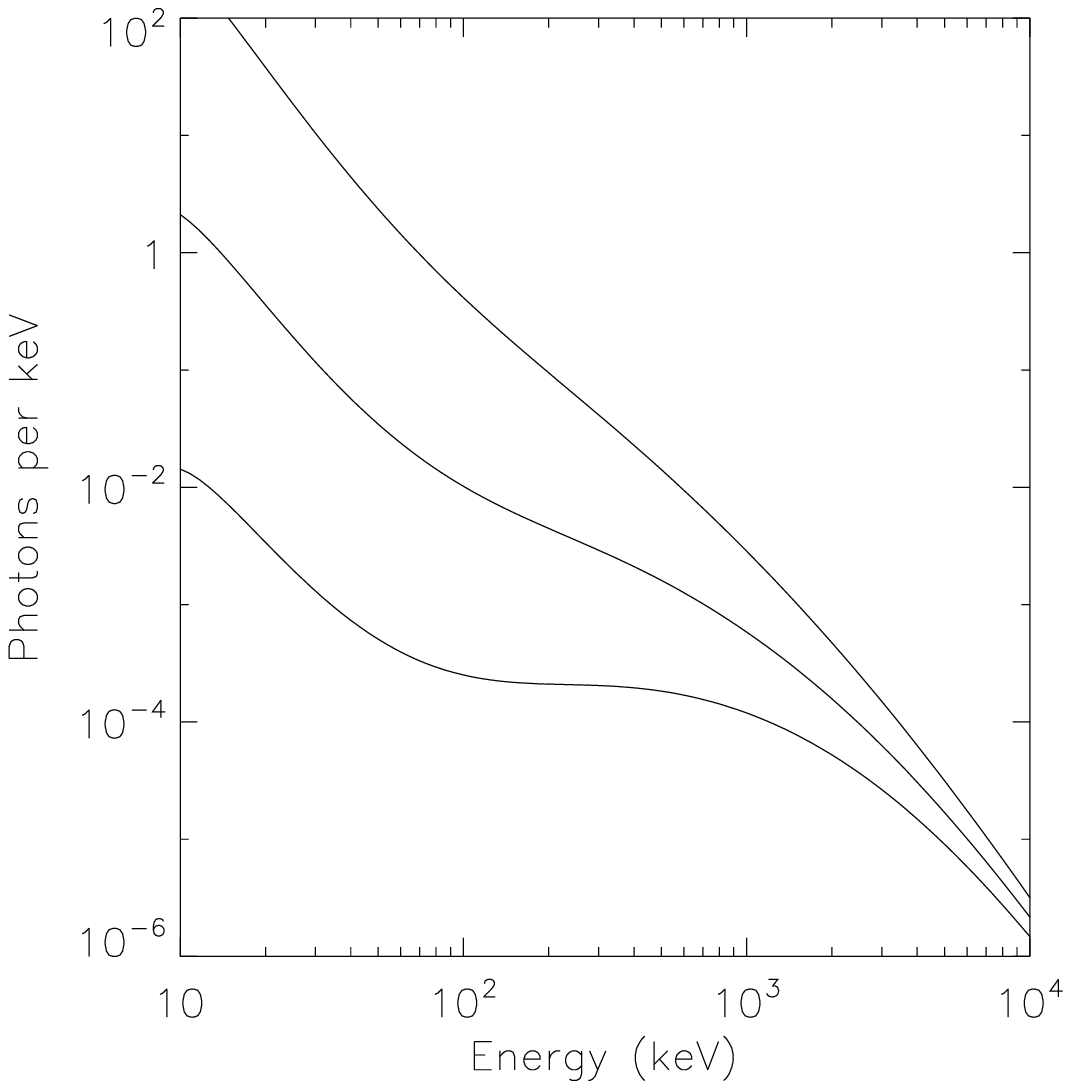}
\hskip 0.25in
\vbox to 4.0in {
\vfill
\begin{minipage}[b]{2.25in}
{\bf Fig. 1}---Compton attenuated spectrum.
For all three curves, the spectral parameter $\delta = -4$.  From top
to bottom, the curves have $\tau_T = 10$, 15, and 20, corresponding to
$\Psi = -0.20$, $-0.13$, and $-0.10$ respectively.
\end{minipage}
\vfill
}}
\end{figure}

This spectrum is defined by four parameters in the gamma-ray energy band:
a Thomson optical depth $\tau_T$, a parameter
$\Psi \equiv \left( \delta + 2 \right)/\tau_T < 0$, where $\delta$ is the
power-law index of the spectrum as $\nu \rightarrow \infty$, a redshift $z$,
and an overall normalization. In the discussion that follows, the spectral
parameter $z$ is set to 0, so that the cosmological redshift enters
explicitly through equation (7).  The overall normalization is folded
into the free parameter $L$, which is a luminosity that is defined in
equation~(6).  The two remaining parameters describe the shape of the
spectrum.  The parameter $\Psi$ sets the peak energy $E_p$ of $\nu F_\nu$
in the cosmological comoving frame, and the parameter $\tau_T$ defines
the width of $\nu F_\nu$ about $E_p$.  The energy $E_p$ only exists for
$\Psi > -0.1972$, taking on a minimum value of $231 \, \keV$ at
$\Psi = -0.1972$.  For $0 < -\Psi \ll 0.1972$, the value of $E_p$ is given
by $\Psi = -\log\left( 2 E_p/m_e c^2 \right) \, m_e c^2/E_p$, so a wide
range of $\Psi$ values produces a narrow range of~$E_p$ values.
The width in $\log E$ of the $\nu F_\nu$ curve about $E_p$ is proportional
to $1/\sqrt{ \tau_T }$.  General values of these parameters that produce
good fits to the data are $10 \lesssim \tau_T \lesssim 30$ and
$-0.16 < \Psi < -0.12$
(\markcite{Brainerd5}Brainerd \etal\ 1996b).
Three examples are given in Figure~1.  In this figure, $\delta=-4$ and
$\tau_T = 10$, 15, and 20, which gives $\Psi = -0.20$, $-0.133$,
and~$-0.10$ respectively; the uppermost curve does not have an $E_p$
since $\Psi < -0.1972$.  The inflection point at $231 \, \keV$
is apparent in these spectra.  The x-ray upturn predicted by
this spectrum has been observed
(\markcite{Preece}Preece \etal\ 1996;
\markcite{Brainerd4}\markcite{Brainerd5}Brainerd \etal\ 1996a,b).
In this study, only the spectrum above $50\, \keV$ is used.


\section{The General Gamma-Ray Burst Flux Distribution}

The equation for the flux distribution of cosmological gamma-ray bursts
is well known.  The general equations for the number of gamma-ray
bursts per unit proper time $\tau$ per unit proper volume $V$ per unit
luminosity $L$ per unit spectral parameter space ${\vec p}$ is
\begin{equation}
	{ d N \over d \tau \; d V \; d L \; d^n p }
   =\rho\left( \tau, r, L, {\vec p} \, \right)
   = R\left( \tau \right)^{-3} \,
	n_0\left(\tau\right) \, \Phi\left( L, {\vec p} \, \right) \, ,
\end{equation}
where the second equality expresses the assumption that the characteristics
of a gamma-ray burst are independent of the age of the universe.
The function $n_0\left(\tau\right)$ is the number of gamma-ray bursts per
unit proper time per unit comoving volume.  The function
$\Phi\left( L, {\vec p}\, \right)$ is the fraction of gamma-ray bursts
per unit luminosity per unit spectral parameter space; it is normalized
to unity.  The proper time of the gamma-ray burst source is related
to the observer's time by
\begin{equation}
	dt = \left( 1 + z \right) \; d\tau
   \, ,
\end{equation}
and the volume element is related to the coordinate elements through
\begin{equation}
	d V = R\left( \tau \right)^{3} \,
	{ r^2 \over \sqrt{ 1 - \left( 2 q_0 - 1 \right) r^2 } } \; dr \; d\Omega
   \, ,
\end{equation}
where the nonstandard coordinate radius $r$ is defined in terms of
the Robertson-Walker coordinate radius $r_1$ as $r = r_1 H_0 R_0 /c$.
The parameter $R\left(\tau\right)$ is $c/H_0 R_0$ times
the expansion parameter for the Robertson-Walker metric, and
the factor $\sqrt{ 1 - \left( 2 q_0 - 1 \right) r^2 }$ in equation (3)
is the familiar $\sqrt{ 1 - k r_1^2 }$ from the Robertson-Walker metric.
The coordinate radius is related to the redshift by
\begin{equation}
	r = { q_0 z + \left( q_0 - 1 \right)
   \left( \sqrt{ 1 + 2 q_0 z } - 1 \right)
   \over q_0^2 \left( 1 + z \right) }
   = { 2 z \left(\, 1 + z + \sqrt{ 1 + 2 q_0 z } \, \right)
   \over \left( 1 + z \right) \left(\, 1 + \sqrt{ 1 + 2 q_0 z }\, \right)^2 }
   \, .
\end{equation}
From the formal derivation of equation (4)
(\markcite{Weinberg}Weinberg 1972, eq.~[15.3.23]),
the first derivative of $r$ with $z$ is
\begin{equation}
	{ dr \over dz } = { \sqrt{ 1 - \left( 2 q_0 - 1 \right) r^2 }
   \over \sqrt{ 1 + 2 q_0 z } \, \left( 1 + z \right) }
   \, .
\end{equation}

The gamma-ray number flux $F$ in an energy band spanning the
frequencies $\nu_1$ to $\nu_2$ is related to the luminosity by
\begin{equation}
	F = { L \, H_0^2 \over c^2 } \, {\cal F}\left( z, {\vec p} \, \right)
   \, ,
\end{equation}
where the length scale $R_0$ is folded into the luminosity $L$, and
the function ${\cal F}\left( z, {\vec p} \right)$, which is defined as
\begin{equation}
	{\cal F}\left( z, {\vec p} \right) = r^{-2} \int_{\nu_1}^{\nu_2}
   I \left[ \nu \left( 1 + z \right), {\vec p} \, \right] \; d \nu \, ,
\end{equation}
is the photon number flux normalized by the overall luminosity.
The function $I \left( \nu \right)$ is the photon number flux per
unit frequency.
The purpose of this particular parameterization is to separate
the luminosity $L$ from the spectral shape parameters ${\vec p}$.

With equations (2) through (7), one can transform equation (1)
into a distribution over observable parameters:
\begin{equation}
	{d N \over d t \; d \Omega \; d F \; d z \; d^n p }
   = n_0 \left[ \tau\left(z\right) \right] \, 
   \Phi\left( { F \over {\cal F}\left( z \right) }, {\vec p} \right)
   \, {\cal F}\left( z \right)^{-1} \,
   { r^2 \over \left( 1 + z \right)^2 \, \sqrt{ 1 + 2 q_0 z } } \,
   \, .
\end{equation}
This equation differs from the equation for galaxy distributions
by a factor of $1/\left( 1 + z \right)$, which arises from the time
dilation of the burst rate.

The distribution of bursts with flux is given by the integration of
equation (8) over all~$z$.  For this integral to be bounded,
$\Phi\left( L, {\vec p}\, \right)$ must have limits on its asymptotic
behavior as $L \rightarrow 0$ and $L \rightarrow \infty$. If we assume that
$\Phi\left( L, {\vec p}\, \right) \propto L^{-\beta_l}$ for $L \ll L_0$,
where $L_0$ characterizes the peak of the function
$L^{5/2} \, \Phi\left( L, {\vec p}\, \right)$, then the integral of
equation (8) over $z$ converges as $z \rightarrow 0$ when $\beta_l < 5/2$.
If $\Phi\left( L, {\vec p}\, \right) \propto L^{-\beta_h}$ for $L \gg L_0$,
the integral converges as $z \rightarrow \infty$ when $\beta_h > 5/2$.
While the second limit is derived under the assumption that $z \ll 1$ when
$L \gg L_0$, relaxing this assumption lowers the value on the right side
of the inequality, so the condition holds for all parameter space.

\subsection{Asymptotic Behavior}

The flux appears in equation (8) only in the argument of the luminosity distribution function.  The asymptotic behavior with flux of the integral
of equation (8) over $z$ is therefore determined by the behavior of
$\Phi\left( L, {\vec p}\, \right)$ at the values of $z$ producing the
greatest contribution to the integral.  If this occurs for values of
$z$ that give $L \approx L_0$, then the integration variable $z$ can be
replaced by $L$, and the expansion of factors involving $z$ determine the
asymptotic behavior with $F$.  On the other hand, if the integrand peaks
at values of $z$ such that $L \ll L_0$ or $L \gg L_0$, then it is the
asymptotic behavior of $\Phi\left( L, {\vec p}\, \right)$ that determines
the asymptotic behavior of the flux distribution with~$F$.

When $z \ll 1$ for $L \gg L_0$, the flux is $ \propto L/z^2$, and equation
(8) is $\propto \Phi\left( L, {\vec p}\, \right) z^4$, assuming no density
evolution at $z \ll 1$.  The integrand peaks at $L_0$, so its contribution
to the integral above $z \approx 1$ is negligible.  Integrating over $z$
and changing the variable of integration to $L$ then gives the flux
distribution, which is proportional to $F^{-5/2}$. This is the origin of
the $-3/2$ slope of the $\log N$--$\log F$ curve of a homogeneous distribution.

Of more interest is the behavior of equation (8) when $z \gg 1$ for
$L \approx L_0$.  To discuss the asymptotic behavior of the integral in
this regime,  I let the photon spectrum be a power law, so that
$I_\nu \propto \nu^{\delta}$, with $\delta < -2$.  This is justified
by the observed power-law behavior above $1 \, \MeV$.  I also assume that
$\Phi\left( L, {\vec p}\, \right) \propto L^{-\beta}$ when $L \gg L_0$ or
$L \ll L_0$, with $\beta = \beta_h$ for the first condition and
$\beta = \beta_l$ for the second.  Finally, I set
$n_0\left(\tau\right) \propto \left( 1 + z \right)^{\mu}$ to investigate
the effects of density evolution.  A necessary and sufficient condition
for the integral over equation (8) to be bounded under these assumptions
is that $\beta_l < 5/2$, which guarantees the convergence of the integral
as $z \rightarrow 0$, and $\beta_h > 1 + \left( 3/2 - \mu \right)/\delta$,
which guarantees convergence as $z \rightarrow \infty$.  Because
$\beta_h > 5/2$ is already required for convergence in the regime of
$z_0 \ll 1$, this last criterion is satisfied when
$\mu < 3\left( 1 - \delta \right)/2$.  Because $\delta < -2$, the right
side of this inequality is $> 9/2$, so this second condition on $\beta_h$
is only important for very strong evolution.  I assume that this limit on
$\mu$ holds in the discussion below.

When $z \gg 1$ for $L \approx L_0$, the integral can be divided into
four parts that are bounded by $z = 1$, $z = 1/q_0 > 1$, and
$L\left(z\right) = L_0$.  When the integral peaks at a value $z_0$ that
is either of these first two values, the factor of $F$ in the argument
of the luminosity function can be factored out of the integral.  The
distribution in flux is then proportional to $F^{-\beta}$.  There remains
two questions to answer: when is $z_0$ at $L\left(z_0\right) \approx L_0$,
and what is the dependence on~$F$ of the flux distribution in this case?

Two regimes exist that affect the dependence of the burst distribution
when $L\left(z_0\right) \approx L_0$:  $1 < z_0 < 1/q_0$ and $z_0 > 1/q_0$.
In the first regime, equation (8) has the asymptotic integral form of
\begin{equation}
	{d N \over d t \; d \Omega \; d F \; d^n p }
   = n_c \, \int_0^{\infty}
   \Phi\left( { F \over {\cal F}\left( z \right) }, {\vec p} \right)
   \, {\cal F}\left( z \right)^{-1} \, z^{\mu} \; dz
   \, .
\end{equation}
Because
${\cal F}\left( z \right) \propto z^{-2-\delta}$ in this regime (eq.\ [7]
and [4]), this equation becomes
\begin{equation}
	{d N \over d t \; d \Omega \; d F \; d^n p }
   \propto F^{ -1 - { 1 + \mu \over 2 - \delta } }
   \, .
\end{equation}
Necessary and sufficient conditions for $L\left(z_0\right) \approx L_0$
are $\beta_l < 1 + { 1 + \mu \over 2 - \delta }$
and $\beta_h > 1 + { 1 + \mu \over 2 - \delta }$.  Because I am assuming
$\mu < 9/2$ and $\delta < -2$, the strongest condition on $\beta_h$
is $\beta_h > 19/11$, which is weaker than the earlier condition of
$\beta_h > 5/2$.  As a consequence, one never can have $z_0 \approx 1/q_0$.
The only possibility is to violate the condition on $\beta_l$, in which
case $z_0 \approx 1$ in the $1< z_0 < 1/q_0$ regime.  Because of this,
the general asymptotic equations for the burst flux distribution is
\begin{equation}
	{d N \over d t \; d \Omega \; d F \; d^n p }
   \propto F^{-\alpha_1}
   \, ,
\end{equation}
where
\begin{equation}
	\alpha_1 = \max\left( \beta_l, 1 + { 1 + \mu \over 2 - \delta } \right)
   \, .
\end{equation}

In the second regime, where $z_0 > 1/q_0$, the asymptotic form of the integral
over equation (8) is
\begin{equation}
	{d N \over d t \; d \Omega \; d F \; d^n p }
   = n_c \, \int_0^{\infty}
   \Phi\left( { F \over {\cal F}\left( z \right) }, {\vec p} \right)
   \, {\cal F}\left( z \right)^{-1} \, z^{\mu - {5\over 2}} \; dz
   \, .
\end{equation}
In this regime, ${\cal F}\left( z \right) \propto z^{\delta}$, so
the dependence of the flux distribution on $F$ is 
\begin{equation}
	{d N \over d t \; d \Omega \; d F \; d^n p }
   \propto F^{ -1 + { \mu - 3/2 \over \delta } }
   \, .
\end{equation}
Sufficient (but not necessary) conditions for $L\left(z_0\right) \approx L_0$
are $\beta_l < 1 + { 1 + \mu \over 2 - \delta }$,
$\beta_l < 1 - { \mu - 3/2 \over \delta }$,
and $\mu > \left( 6 - 5 \delta \right)/4$.  The last condition guarantees
that the integration does not have secondary maximum at $z = 1$.
Because I assume
$\delta < -2$ and $\mu < 9/2$, this last condition is always satisfied.
As a consequence, the second condition on $\beta_l$ is the stronger, and
the general asymptotic solution can be written as
\begin{equation}
	{d N \over d t \; d \Omega \; d F \; d^n p }
   \propto F^{-\alpha_2}
   \, ,
\end{equation}
where
\begin{equation}
	\alpha_2 = \max\left( \beta_l,
   1 - { \mu - {3\over 2 } \over \delta } \right)
   \, .
\end{equation}

Several points are immediately clear from these equations. First,
the behavior of the luminosity distribution function below $L_0$
influences the asymptotic behavior of the flux distribution, while
the behavior above $L_0$ is unimportant.  Recall that $L_0$ is defined
as the maximum of $L^{5/2} \Phi\left(L, {\vec p} \, \right)$.
Second, in the absence of strong source evolution,
a large asymptotic power-law index at small $F$ requires a luminosity
distribution with a power-law index of equal value, as noted by
M\'esz\'aros \& M\'esz\'aros \markcite{Meszaros}(1995).
Because $\delta < -2$, the minimum value of $\alpha_1$, which is given
by the second term in equation (12), is $< 5/4$
for $\mu = 0$; from the second term of equation (16),
the minimum value of $\alpha_2$ is $< 1/4$ for $\mu = 0$.
Third, because the luminosity function can formally diverge
and still give a bounded integral as long as it falls below a $L^{-5/2}$
power law as $L \rightarrow 0$, the presence of a change in the
luminosity distribution function at sufficiently low luminosity---which
is required by the boundedness of $\int \Phi\left(L\right) \; dL$---has
no effect on the observed flux distribution; formally, this must be
at $L \ll F_{min} c^2/H_0^2 {\cal F} \left(1\right)$, where
$F_{min}$ is the minimum observable burst flux
(\markcite{Meszaros}M\'esz\'aros \& M\'esz\'aros 1995).
Finally, evolution is unimportant in the asymptotic behavior unless
the number of bursts increases dramatically as $1 + z$ increases;
the second index in equation (12) increases by $< \mu/4$, and the
second index in equation (16) increases by $< \mu/2$, so evolution
cannot dominate the luminosity distribution function unless $\mu \gtrsim 3$.

Equations (12) and (16) imply different behaviors for
$q_0 = 0.5$ and $q_0 \ll 0.5$.  For $q_0 = 0.5$, the only relevant limit
is that given by equation (16).  When $z_0 \ll 1$, the integrand rises
with $z$ until the peak in $\Phi\left(L,{\vec p}\,\right)$
is reached.  When $z_0 \gg 1$, the integrand rises until $z \approx 1$;
then, if $\beta_l > 1 - \left(  \mu - {3\over 2 } \right)/\delta$,
the integrand goes to zero more rapidly than $z^{-1}$ as
$\rightarrow \infty$, but if the inequality is violated,
the integrand either rises or goes to zero less rapidly than $z^{-1}$.
In contrast, when $q_0 \ll 0.5$, there are three regimes to consider.
As in the previous case, when $z_0 < 1$, the integrand rises to the
maximum of $\Phi\left(L,{\vec p}\,\right)$,
but for $1 < z_0 < 1/q_0$, equation (12) applies, and
the integrand goes to zero more rapidly than $z^{-1}$ when
$\beta_l > 1 + \left( 1 + \mu \right)/\left( 2 - \delta \right)$, and
rises or goes to zero less rapidly than $z^{-1}$ when the inequality
does not hold.  When $z_0 > 1/q_0$, equation (16) holds, and the
behavior noted for $q_0 = 0.5$ holds for $q_0 \ll 0.5$.
The consequences of this behavior are that the minimum value of $\alpha$ in
an $F^{-\alpha}$ asymptotic description of the flux distribution
is strongly dependent on $q_0$ and that, for $q_0 \ll 0.5$,
the distribution in $z$ for bursts with a given $F$ is broad and flat over
$1 < z_0 < 1/q_0$ when
$1 + \left( 1 + \mu \right)/\left( 2 - \delta \right) >
\beta_l > 1 - \left(  \mu - {3\over 2 } \right)/\delta$.


\section{Incorporating a Luminosity Distribution}

The gamma-ray burst flux distribution is derived for two luminosity
distributions: a monoluminous distribution function, and a
power-law luminosity distribution with a high luminosity cutoff.  The
first is widely used, and it is given here for comparison to the
power law distribution.

\subsection{Monoluminosity Distribution}

The equation for a monoluminous gamma-ray burst distribution does not have
an integral over $z$. It is therefore relatively easy to evaluate and
fit to the observed flux distribution.  From equations (1) through (6),
\begin{equation}
	{ d N \over d F } = N_0 { 4 r^2 \, \sqrt{ 1 + 2 q_0 }
   \over \left( 1 + z \right)^2 \, r_0^2 \, \sqrt{ 1 + 2 q_0 z } }
   \left[ {\left. { d {\cal F} \left( z \right) / dz } \right|_{z=1}
   \over { d {\cal F} \left( z \right) / dz } } \right] 
   \, .
\end{equation}
In this equation, $r_0$ is the value of $r$ from equation (4) for
$z = 1$.  The normalization $N_0$ is the number of bursts per unit $F$
at $z = 1$.  $F$ is related to $z$ through
\begin{equation}
	F = F_0 { {\cal F} \left( z \right) \over {\cal F} \left( 1 \right) }
   \, ,
\end{equation}
where ${\cal F} \left( z \right)$ is given by equation~(7).

The distributions produced by equation (17) are plotted as dotted lines
in Figure~2 for $F_0 = 1$, $N_0 = 1$, $\Psi = -0.15$, and $\tau_T = 15$
for both $q_0 = 0.5$ and $0.1$.  The curves for $z$ given by equation (18)
as a function of $F$ are plotted in Figure~4 for these parameters.  The
curves in Figure~2 go to the correct limiting power-law indices: for
$q_0 = 0.5$, it is given by equation (16) as $11/17$ ($\approx 0.65$),
while for $q_0 = 0.1$, it is given by equation (12)
as $29/25$ ($\approx 1.16$).

\begin{figure}[t]
\hbox to \hsize{
\epsfysize=3.0in \epsfbox{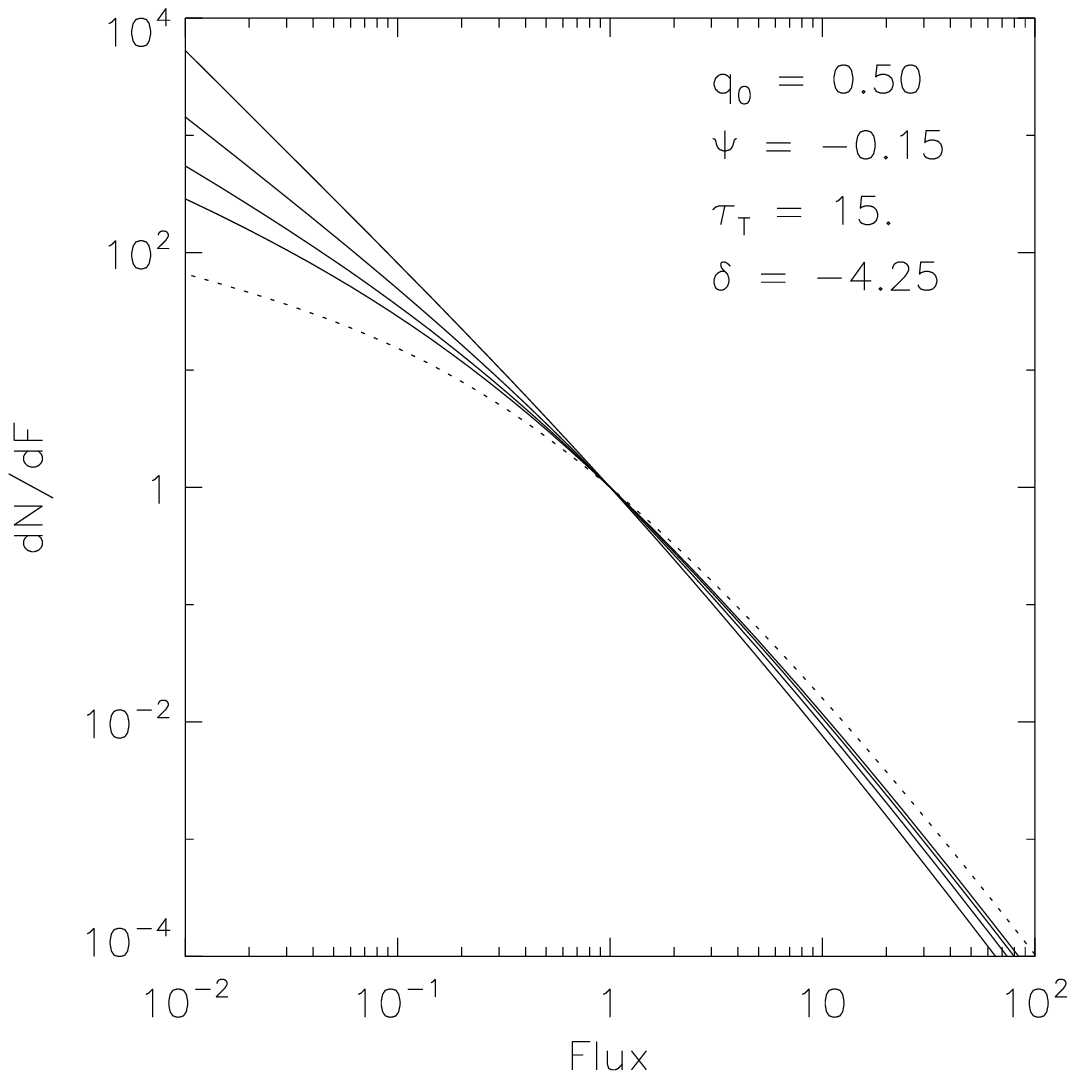}
\hfill
\epsfysize=3.0in \epsfbox{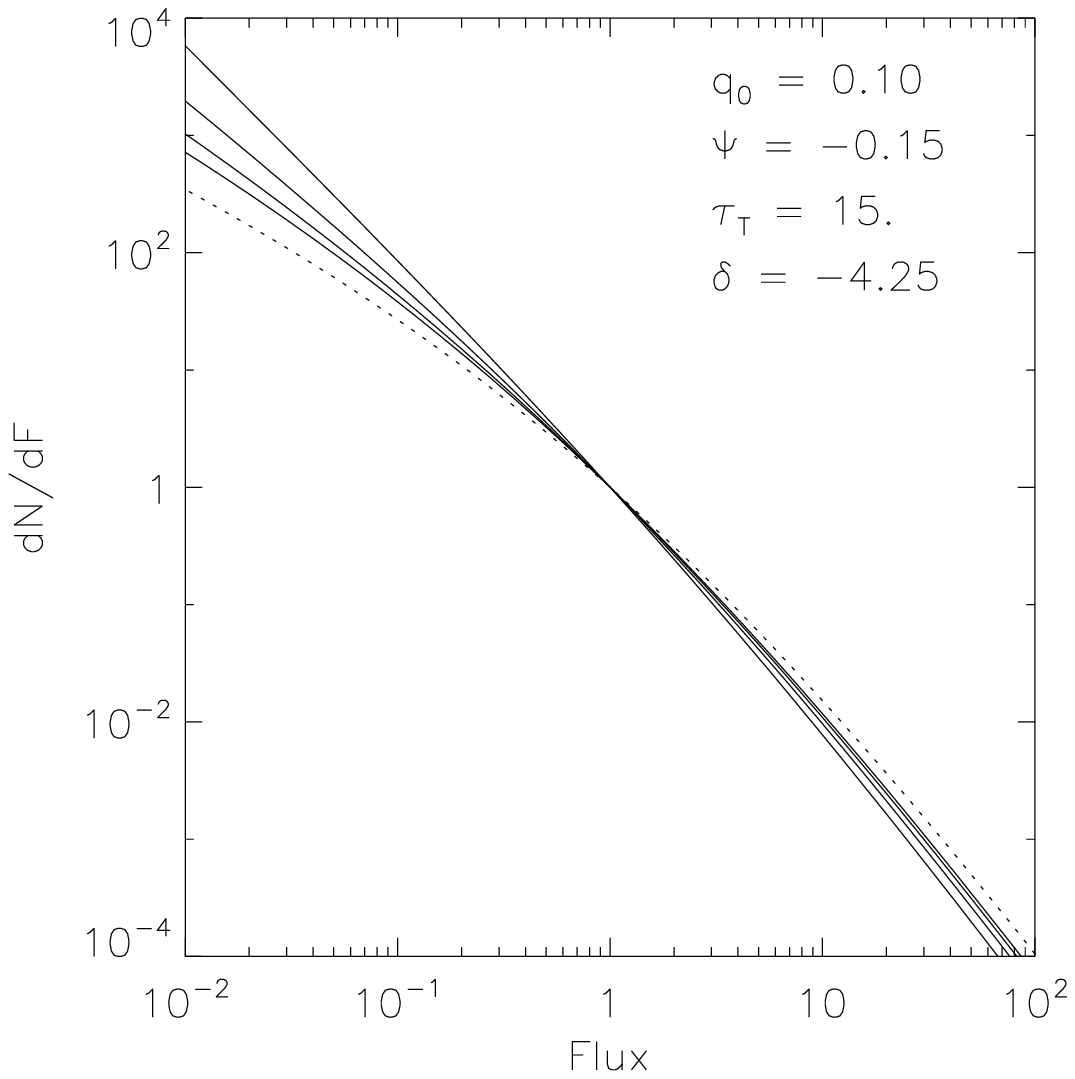}
}
\vskip 0.1in
\hbox to 5.5in { \hskip 0.5in
\begin{minipage}[b]{5.5in}
{\bf Fig. 2}---Peak-flux distributions.  All curves are for spectral
parameters $\Psi = -0.15$ and $\tau_T = 15$, which are defined in \S 2.
The figures are for $q_0 = 0.5$ (left) and $0.1$ (right).  The solid
curves for a power-law luminosity distributions with (top left to
bottom left) $\beta = 1.8$, $1.4$, $1.0$, and $0.6$.  The solid curve
is the monoluminosity distribution for the given spectral and
cosmological parameters.  The curves are normalized by $N_0 = 1$
and $F_0 = 1$ so that $d N /d F = 1$ and $z_{max} = 1$ at $F = 1$.
The flux is measured in the $50 \, \keV$ to $300 \, \keV$ energy band.
\end{minipage}
\hfill
}
\end{figure}

\subsection{Cutoff Power-Law Luminosity Distribution}

A simple distribution function used frequently in gamma-ray burst studies
is a power-law distribution with one or two luminosity cutoffs.  In this
study, I use
\begin{equation}
	\Phi\left(L\right) = \cases { \left( { L \over L_0 } \right)^{-\beta} \, ,
   &for $L \le L_0$; \cr
   \noalign{\vskip 10pt}
   0 \, , &otherwise. \cr }
\end{equation}
Placing this into equation (8) and integrating over $z$ gives

\begin{equation}
	{ d N \over d F }
   = n_0 \left( {F \over F_0 }\right)^{-\beta} \int_0^{z_{max}}
   {\cal F}^{\beta - 1}\left( z \right)
   { r^2 \over \left( 1 + z \right)^2 \sqrt{ 1 + 2 q_0 z } } \; dz
   \, ,
\end{equation}
where $z_{max}$ is defined for
a given flux $F$ by equation (18) with $z$ replaced
by $z_{max}$.  The normalizing flux $F_0$ is the flux at which
$z_{max} = 1$.  The normalization parameter $n_0$ is defined as
\begin{equation}
	n_0 = N_0/\int_0^{1}
   {\cal F}^{\beta - 1}\left( z \right)
   { r^2 \over \left( 1 + z \right)^2 \sqrt{ 1 + 2 q_0 z } } \; dz
   \, ,
\end{equation}
where $N_0$ is the number of gamma-ray bursts per unit flux at $F = F_0$.

Flux distributions from equation (20) with $F_0 = 1$ and $N_0 = 1$
are plotted in Figure~2 as solid lines for $\beta = 0.6$, $1.0$, $1.4$,
and $1.8$ for both $q_0 = 0.5$ and $0.1$.  The asymptotic behavior
discussed in \S 3.1 is clearly demonstrated in these figures.
For $q_0 = 0.5$, equation (16) states that the asymptotic power-law index
is given by $-\beta$ when $\beta > 11/17 \approx 0.65$.  Three of the
four solid curves in this plot satisfy this criterion, and all three
are close to their asymptotic values at low flux.  The fourth,
which has $\beta = 0.6$, is slowly going to the asymptotic value
of $-\beta = -11/17$.  For $q_0 = 0.1$, the limit on $\beta$ is given
by equation (12), because $z_{max} > 6$ for the fluxes in the plot,
and equation (16) is not applicable until $z_0 \gg 10$.  In this case,
the asymptotic limit is given by $-\beta$ when $\beta > 29/25 = 1.16$.
In Figure~2b, the upper two solid curves go to a power-law index of
$-\beta$ as $F \rightarrow 0$, and the lower two solid curves approach
the monoluminous curve.


\subsection{Average Redshift}

An important difference between a monoluminous distribution and a broad
luminosity distribution is that in the former the redshift and the flux
have a one-to-one correspondence, while in the latter there is a
distribution of redshift values for each flux value.  From equation (20),
one sees that the integrand is independent of $F$, and that the dependence
on flux of the integral arises from the dependence of $z_{max}$ on flux
in equation (18).
The shape of the redshift distribution is therefore independent of flux,
and only the upper limit on $z$ changes with flux. When the integrand
rises monotonically with $z$ for all $z$, the median value of $z$ is
a strong function of $z_{max}$, but when the integrand falls sufficiently
rapidly above some value of $z$, the median value of $z$ goes to a
constant as $z_{max}$ goes to infinity.  From the discussion of the
asymptotic behavior of equation (8), we know that the first
instance occurs when $L\left(z_0\right) \approx L_0$, in which case
the dependence of the burst number distribution on $F$ is determined
by the second index in either equation (12) or equation (16).  In this case,
the asymptotic behavior of the flux distribution is independent of the
luminosity distribution function.  The second instance arises when
$z_0 \approx 1$ or $z_0 \approx 1/q_0$, in which case the asymptotic
behavior of the flux distribution is determined by the luminosity
distribution.

\begin{figure}[t]
\hbox to \hsize{
\epsfysize=3.0in \epsfbox{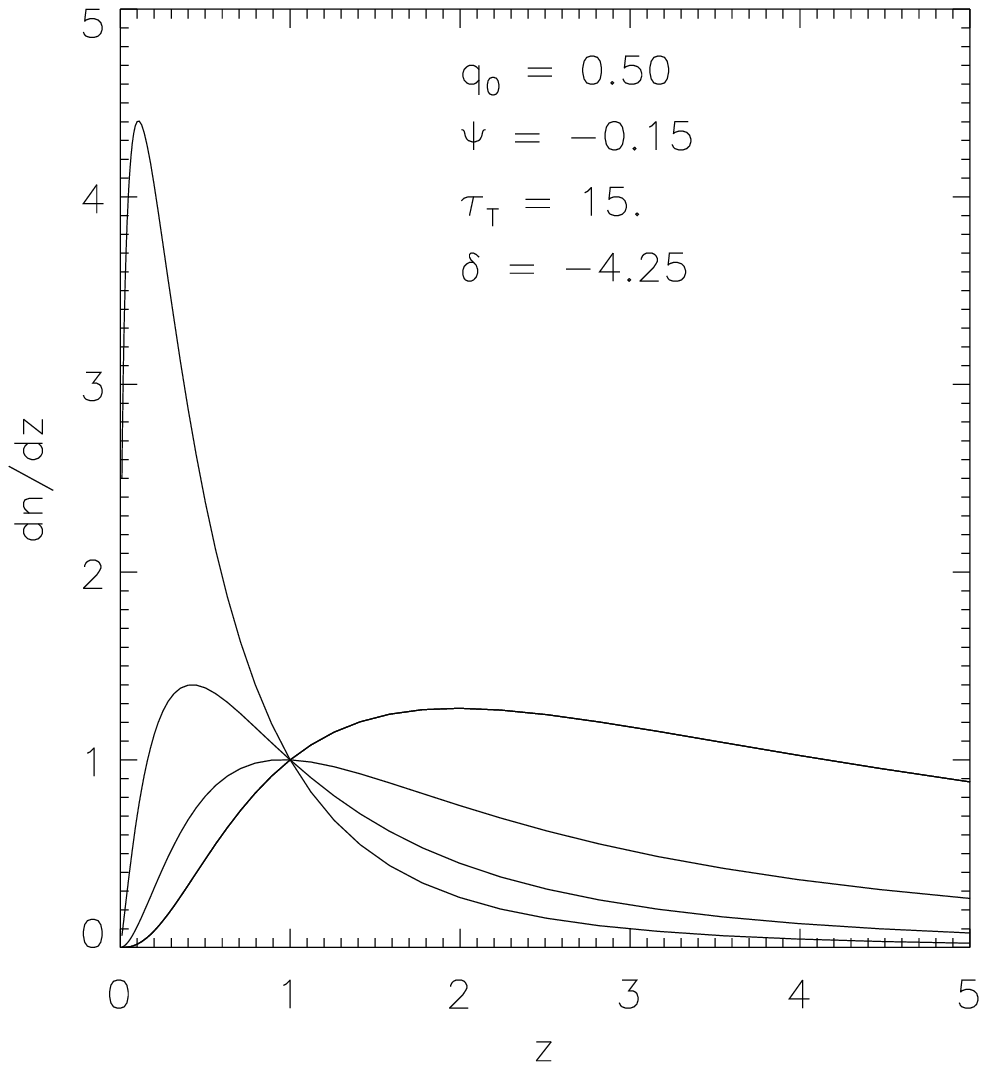}
\hfill
\epsfysize=3.0in \epsfbox{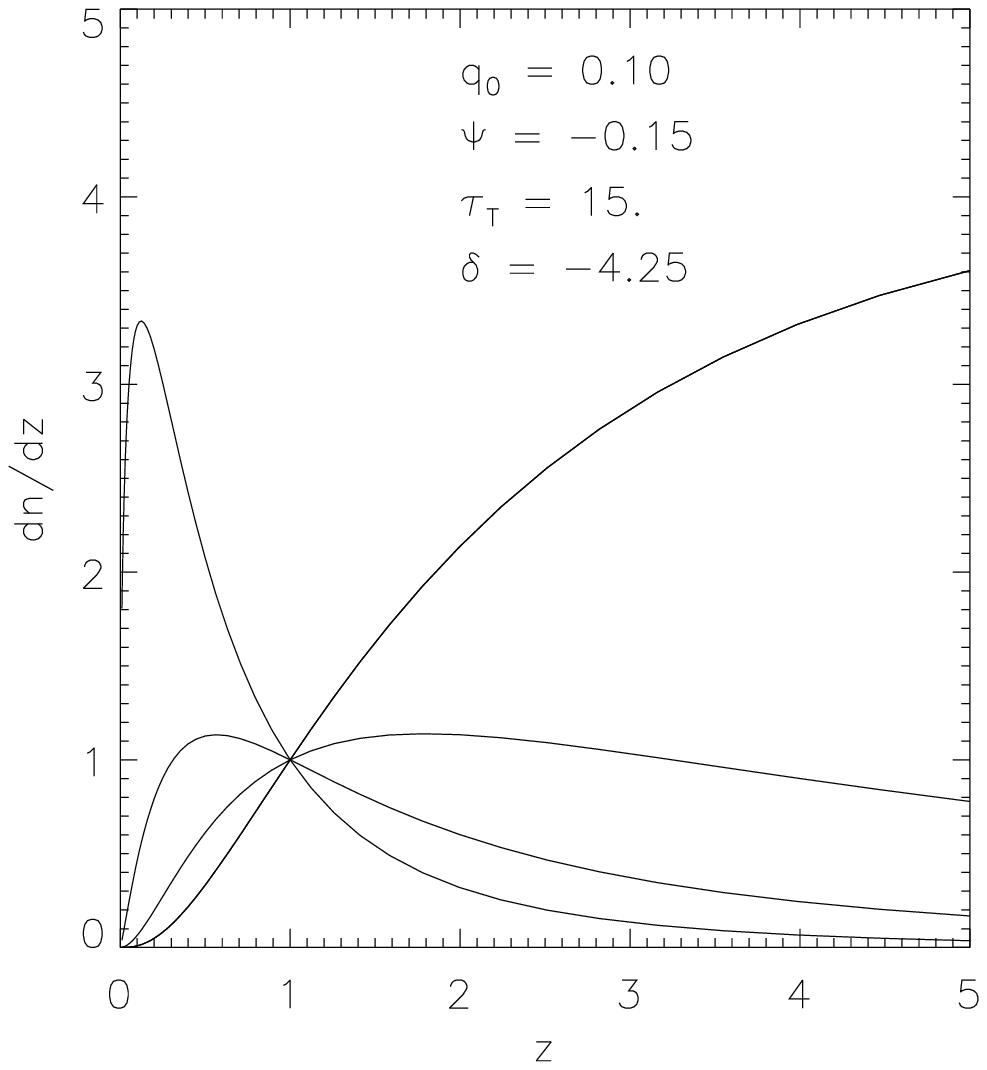}
}
\vskip 0.1in
\hbox to 5.5in { \hskip 0.5in
\begin{minipage}[b]{5.5in}
{\bf Fig. 3}---Distribution
of $z$ for a power-law luminosity distribution.
On each figure, the distribution curves for power-law indexes (from
upper left to lower left) of $\beta = 1.8$, 1.4, $1.0$, and $0.6$
are plotted as functions of cosmological redshift~$z$.  The curves are
normalized to unity at $z = 1$.  The figures are for $q_0 = 0.5$
(left) and~0.1 (right).  The spectral parameters are $\Psi = -0.15$
and $\tau_T = 15$.
\end{minipage}
\hfill
}
\end{figure}

\begin{figure}[t]
\hbox to \hsize{
\epsfysize=3.0in \epsfbox{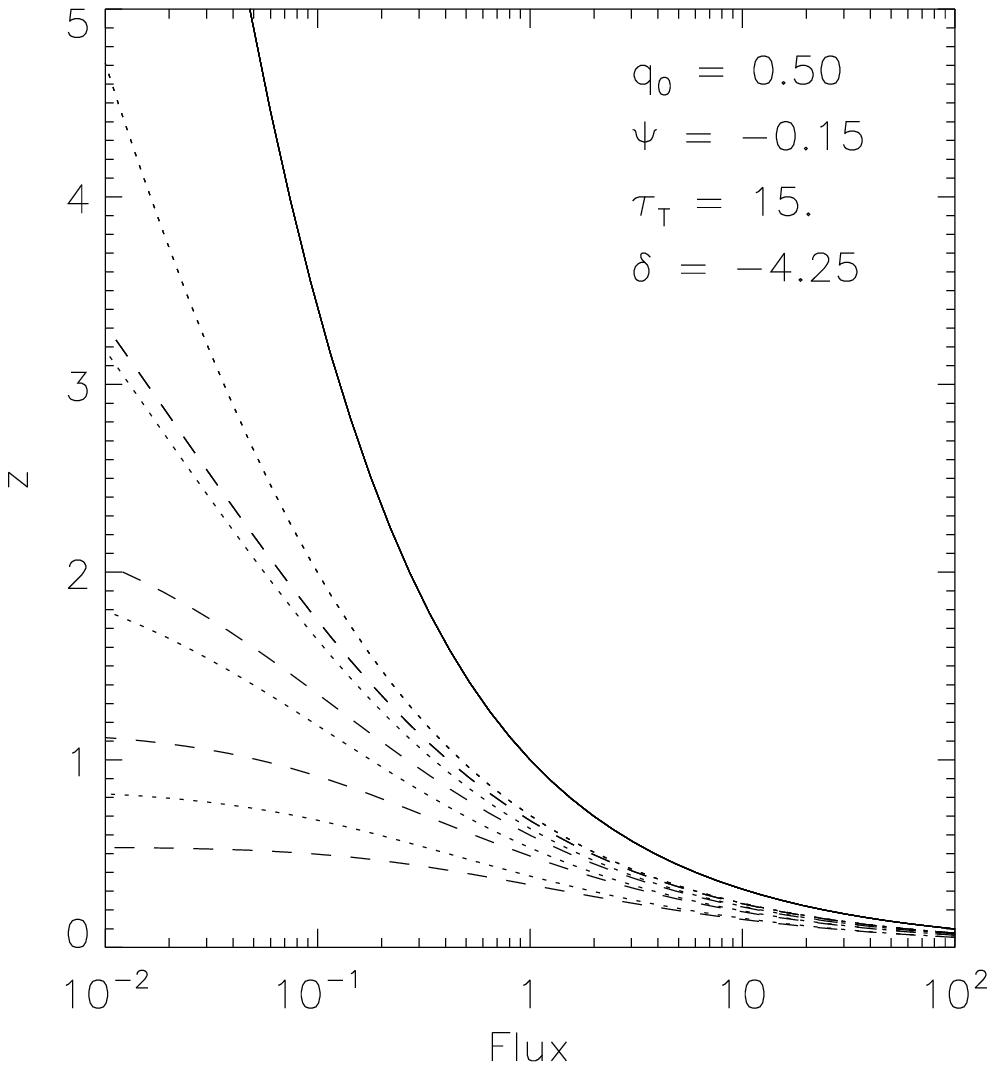}
\hfill
\epsfysize=3.0in \epsfbox{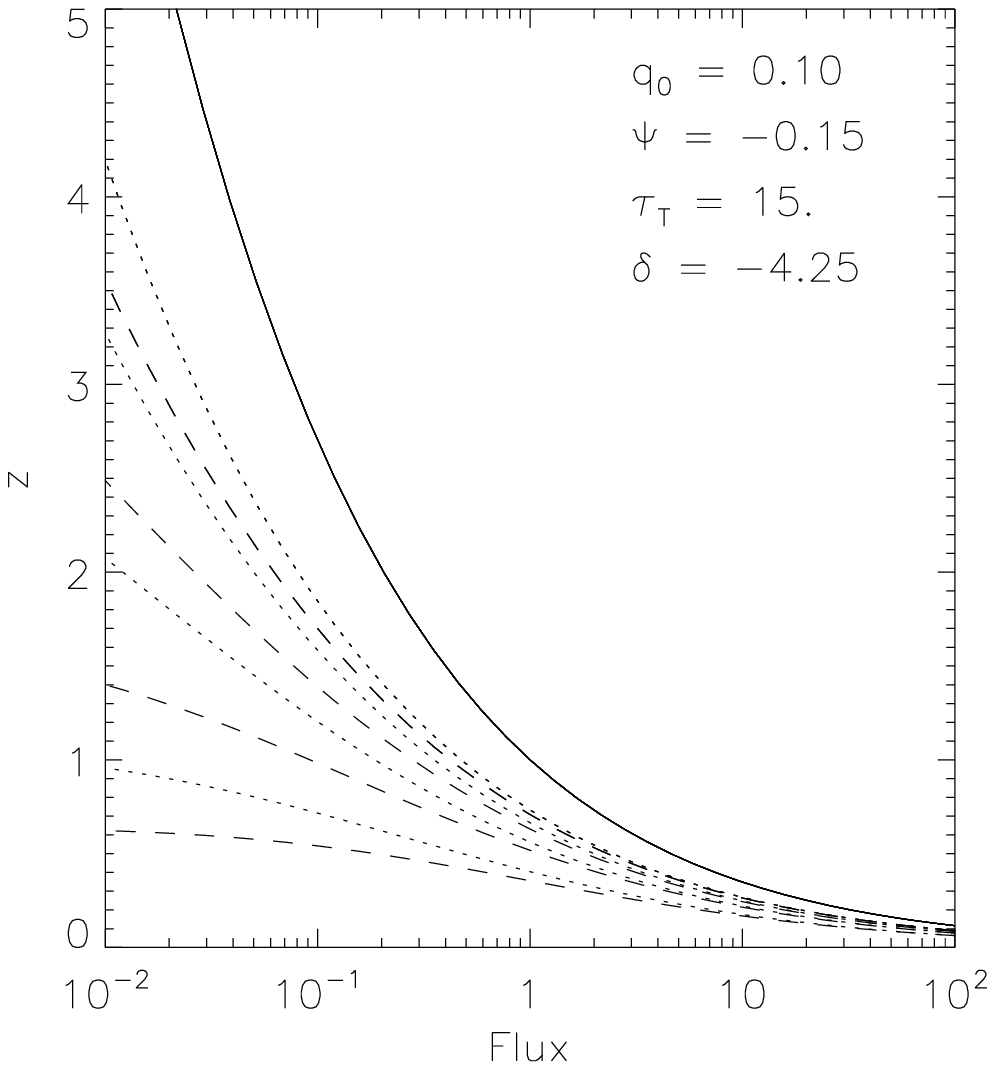}
}
\vskip 0.1in
\hbox to 5.5in { \hskip 0.5in
\begin{minipage}[b]{5.5in}
{\bf Fig. 4}---Average values
of $z$ as a function of flux.  The solid line
gives the relationship between $z$ and $F$ from equation~(18) with
$F_0 = 1$.  The dotted lines give $\left<z\right>$, while the dashed
lines give $1/\left< \left( 1 + z \right)^{-1} \right> - 1$.  The different
lines are for, from bottom to top, $\beta = 1.8$, 1.4, $1.0$, and $0.6$.
The figures are for $q_0 = 0.5$ (left) and~0.1 (right).  The spectral
parameters are $\Psi = -0.15$ and $\tau_T = 15$.
\end{minipage}
\hfill
}
\end{figure}

Figure~3 shows the $z$ distribution function for $\beta = 0.6$, 1.0,
1.4, and 1.8.  In the cases where $\beta$ is large, the distribution
peaks below $z = 1$ and falls rapidly as $z$ increases, so the average
value of $z$ is a weak function of $z_{max}$ when $z_{max} \gg 1$.  As
$\beta$ decreases, the peak in the distribution function moves to higher
$z$, until it disappears altogether, and the distribution function
rises monotonically with $z$, so that the average value of $z$ goes
to $z_{max}$.  Comparing Figure~3a to Figure~3b, one sees the behavior
described in \S 3.1.  When $q_0 = 0.5$, the integrand, and therefore the
distribution in $z$, falls faster than $z^{-1}$ as $z \rightarrow \infty$
when $\beta > 11/17$.  The three curves that satisfy this criterion
exhibit the expected behavior.  The curve that violates the inequality
falls slowly as $z$ rises.  When $q_0 = 0.1$, the criterion is
$\beta > 29/25$, which only two of the curves satisfy.  Of the two curves
violating the criterion, the $\beta = 1$ curve falls slowly, while the
$\beta = 0.6$ curve continues to rise.  When $z > 10$, which is not
shown in the figure, both of these curves fall, since the
$\beta > 11/17$ inequality from equation (12) become the governing
criterion.

Tests for the effects of the cosmological expansion rely on the average
properties of gamma-ray bursts.  Because the bursts of a given peak flux
have a distribution of redshifts, a distinction must be made between
the value of $\left<z\right>$ derived from time-dilation and the
value derived from redshift.  For the first, one is averaging $1 + z$,
while for the second, one is averaging $1/\left( 1 + z \right)$.  The
consequence is that the average value of $z$ derived in the second case 
is smaller than in the first.  This is shown in Figure~4, where
$\left<z\right>$ is plotted as dotted lines, and
$1/\left< \left( 1 + z \right)^{-1} \right> - 1$ is plotted as dashed
lines for the values of $\beta$ given in Figure~2.  The curves go to a
constant as $z \rightarrow \infty$ when they satisfy the criterion on
$\beta$ that makes $-\beta$ the asymptotic power-law index of the flux
distribution.


\section{Fitting the BATSE Flux Distribution}

I use $\chi^2$ minimization to fit the model flux distributions given in
\S 4 to the flux distribution for a subset of gamma-ray bursts from the
BATSE 3B catalog.  This catalog gives the peak fluxes in the energy
range of $50 \, \keV$ to $300 \, \keV$ for 869 gamma-ray
bursts on three different time scales: $64 \, \ms$, $256 \, \ms$, and
$1024 \, \ms$.  The subset I use comprises the bursts that are not
overwrites of a preceding burst, that have a T90 duration
$\ge 2 \, \s$ (where T90 is the time over which 90 \% of the burst
energy is emitted),
and that have a peak flux on the $256 \, \ms$
timescale of $0.7 \, \cm^{-2} \, \s^{-1} < F < 110.9 \, \cm^{-2} \, \s^{-1}$.
There are 396 gamma-ray bursts that meet these criteria.

The upper threshold is for computational convenience, and it has no effect
on the model fits.  The lower threshold is to avoid the effects of the
variable detector threshold and the detector geometry.
The BATSE instrument triggers
on a gamma-ray burst if the count rate exceeds a threshold count rate in
at least two detectors.  This count rate is set by integrating the
background for $17\, \s$ intervals and calculating from this the
standard deviation for counts on each of the three timescales given above.
The trigger threshold is a multiple of this standard deviation; usually
5.5 standard deviations is chosen.  Because the background fluctuates
by a factor of 2 over the 90 minute orbit, the trigger threshold is variable.
Because the trigger is in counts rather than in flux, because the detector
geometry is an octahedron, and because Earth scatters burst radiation to
the experiment, the flux threshold is angle dependent.  These various effects
cause an undercounting of gamma-ray bursts of low flux.  The choice of
$0.7 \, \cm^{-2} \, \s^{-1}$ on the $256 \, \ms$ timescale as a lower limit
overcomes these difficulties
(\markcite{Pendleton1}Pendleton \etal\ 1996).

For 90 minutes after a trigger, the threshold is raised so that if a
particularly bright burst occurs, it is observed by the instrument.
Such bursts, which are called overwrites, must be dropped from the sample,
because they represent an overcounting of bright bursts relative
to dim bursts.

The choice of timescales affects the choice of cosmological model.  The
reason is that if the burst pulse carrying the peak flux is fully resolved
in time, then the cosmological redshift comes into the peak flux through
the Lorentz boost of the flux, but if the pulse is much shorter than the
instrument timescale, all photons in the pulse arrive inside that timescale
regardless of time dilation.  The time dilation of the emission therefore
drops out of the problem, and the peak-flux has one more factor of
$\left( 1 + z \right)$ than it does for the time-resolved
pulse.  I desire a timescale that resolves the time-behavior of most gamma-ray
bursts without introducing large errors in the flux measurement from counting
statistics.
I compare in Figure~5 the peak fluxes of the 396 bursts in my sample.
If two time scales each resolve the time-dependent variation of the flux,
the peak flux measured on each should be equal within counting statistics.
In Figure~5a, each burst is plotted on the $256 \, \ms$--$1024 \, \ms$
timescale plane, and in Figure~5b, each is plotted on the
$64 \, \ms$--$256 \, \ms$ plane.  The upper diagonal line represents equal
fluxes on both timescales, which occurs when a pulse is fully time resolved.
The lower diagonal line represents the short timescale having a peak flux
that is 4 times the peak flux
for the long timescale, which is what occurs when the
pulse is much shorter than the short timescale.  One sees that the clustering
along the line of equality is stronger for Figure~5b than for Figure~5a,
showing that the $256 \, \ms$ timescale resolves most burst while the
$1024 \, \ms$ timescale leaves a large fraction of bursts unresolved.
The percentage of bursts in Figure~5a that have a $256 \, \ms$ peak
flux that is more than a factor of 2 larger than the $1024 \, \ms$ peak
flux is 2.6\%, while the percentage in Figure~5b that have a
$64 \, \ms$ peak flux that is more than a factor of 2 larger than the
$256 \, \ms$ peak flux is 0.7\%.  For a factor of $\sqrt{2}$, these
percentages are 19.8\% and 10.2\% respectively.
This is why the $256 \, \ms$ timescale is used.

\placefigure{fig5}

\begin{figure}[t]
\hbox to \hsize{
\epsfysize=3.0in \epsfbox{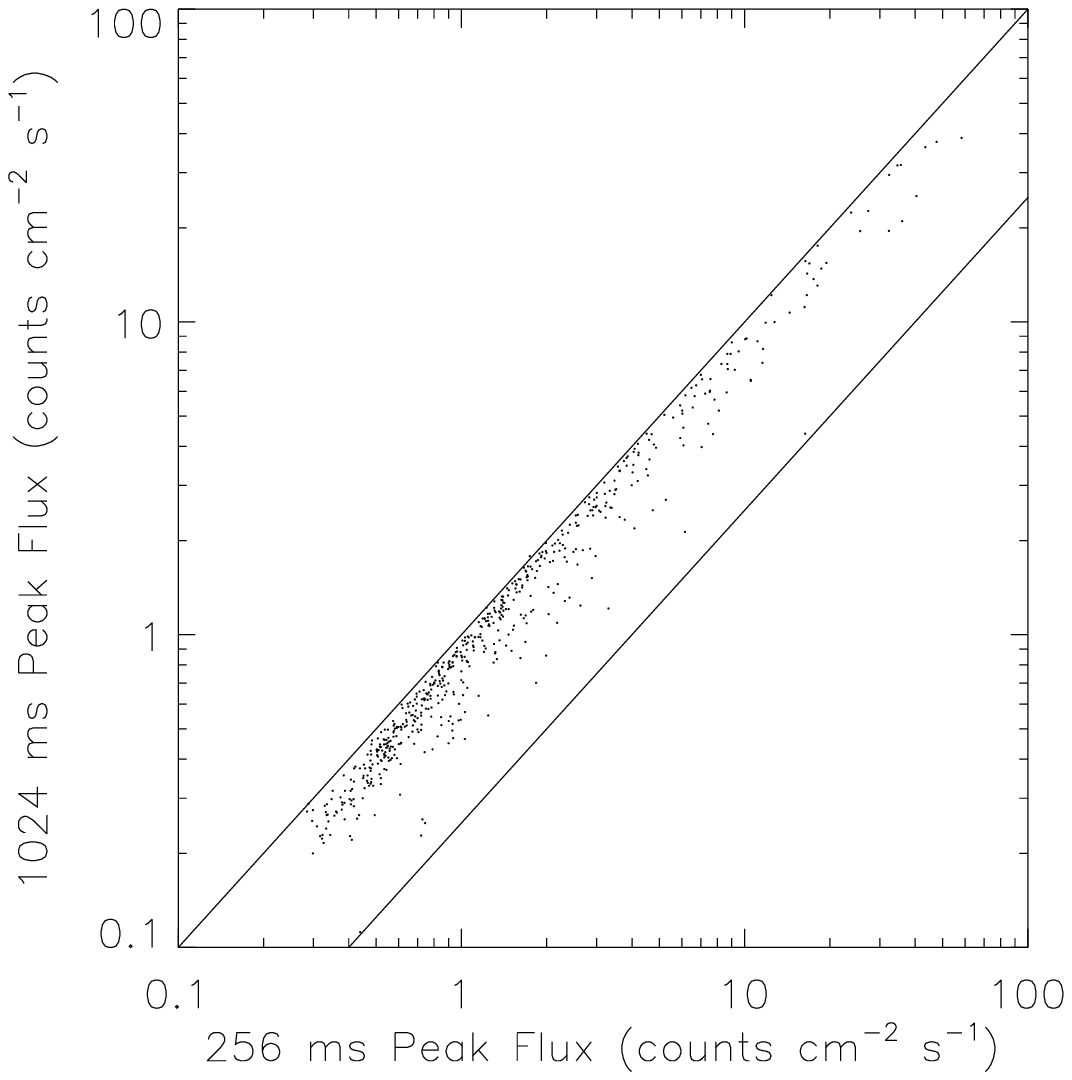}
\hfill
\epsfysize=3.0in \epsfbox{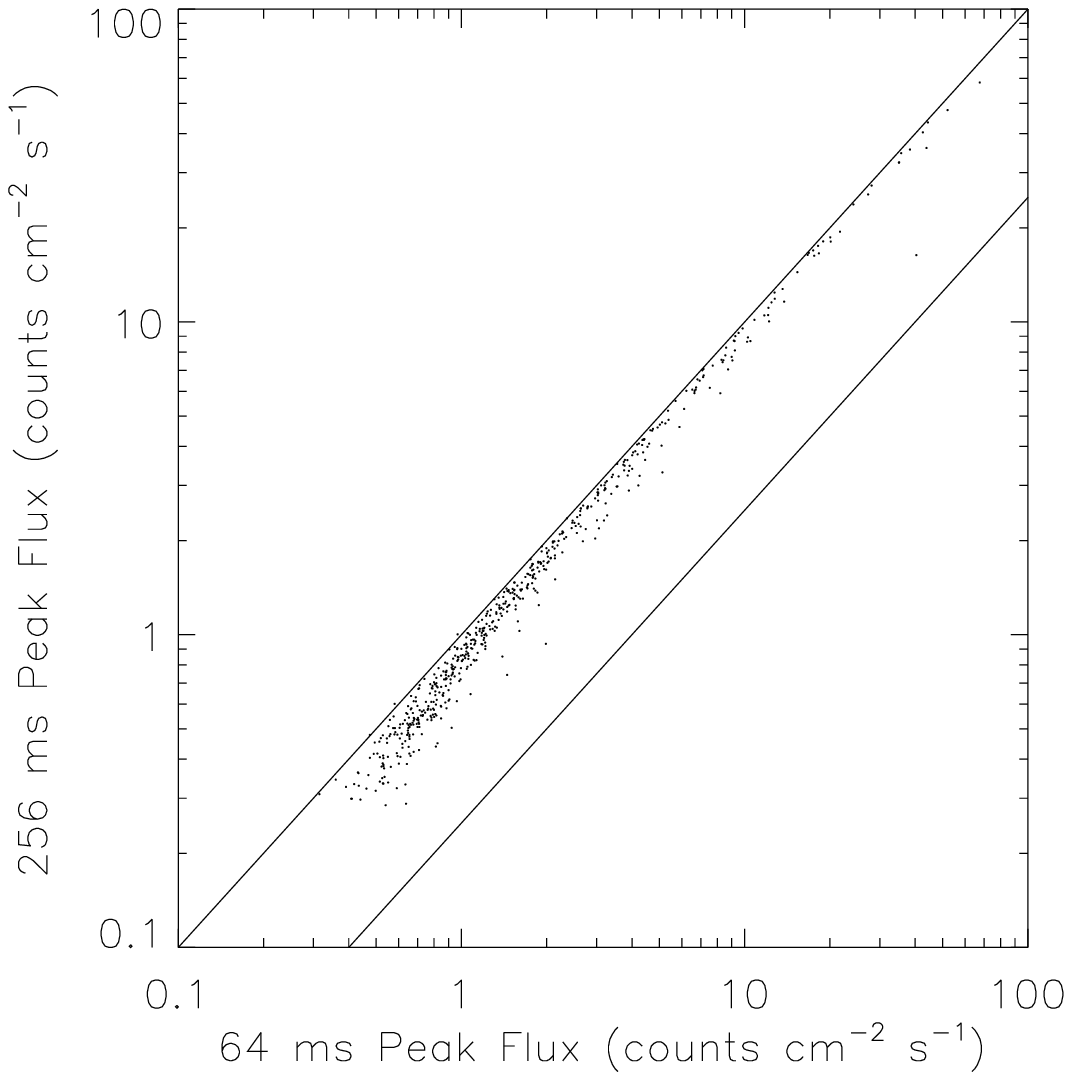}
}
\vskip 0.1in
\hbox to 5.5in { \hskip 0.5in
\begin{minipage}[b]{5.5in}
{\bf Fig. 5}---Peak flux scatter plot.
The gamma-ray bursts from the BATSE 3B catalog that have a duration
measure T90 $> 2 \, \s$ are plotted by their (left) 256 and $1024 \, \ms$
peak fluxes and their (right) 64 and $256 \, \ms$ peak fluxes in the
$50 \, \keV$ to $300 \, \keV$ energy band. The upper diagonal line
on each plot gives flux equality, while the lower diagonal on each
gives a ratio of 1 to 4 in the two peak fluxes.
\end{minipage}
\hfill
}
\end{figure}

Some bursts have durations that are of order the sampling timescale.
For these bursts, the time structure cannot be resolved,
so a lower limit on the gamma-ray burst duration is warranted.  The
duration distribution is bimodal, with a local minimum at $2 \, \s$.
This value is therefore a reasonable lower limit on T90 in the sample.

The burst distribution is summed into 22 bins that uniformly spanning
$\log F$ between $0.7 \, \cm^{-2} \, \s^{-1}$ and $110.9 \, \cm^{-2} \, \s^{-1}$.
Then, going from the highest flux bin to the lowest flux bin, each bin with
fewer than 20 bursts is added to the bin immediately lower in flux.  The
purpose of this is to make gaussian statistics the correct description of
the number of bursts in each bin.  This lowers the number of bins to~11.


\subsection{Monoluminosity Distribution}

\placefigure{fig6}

\begin{figure}[tbp]
\hbox to 6.5in {\epsfxsize=4.0in \epsfbox{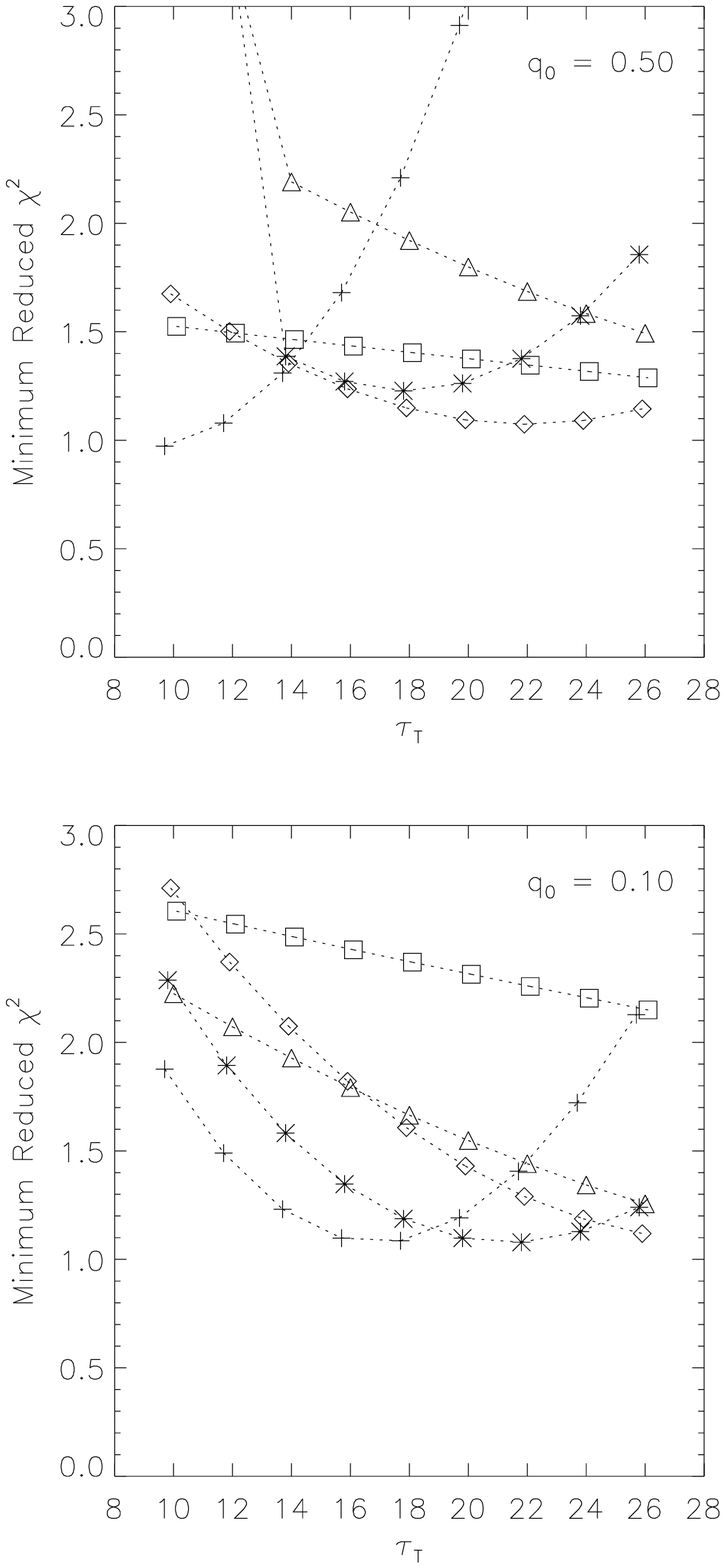}
\hskip 0.25in
\vbox to 8.0in {
\vfill
\begin{minipage}[b]{2.25in}
{\bf Fig. 6}---Reduced $\chi^2$ for fits of monoluminous peak-flux
distributions to the BATSE data.  The x-axis is the spectral parameter
$\tau_T$, and the various sets of points are for
$\Psi = -0.10$ ($+$), $-0.12$ ($\star$), $-0.14$ ($\Diamond$),
$-0.16$ ($\triangle$), and $-0.18$ ($\Box$).  In the upper figure,
$q_0 = 0.5$, and in the lower figure, $q_0 = 0.1$.
\end{minipage}
\vfill
}}
\end{figure}

\begin{figure}[tbp]
\hbox to \hsize{\hfill\epsfbox{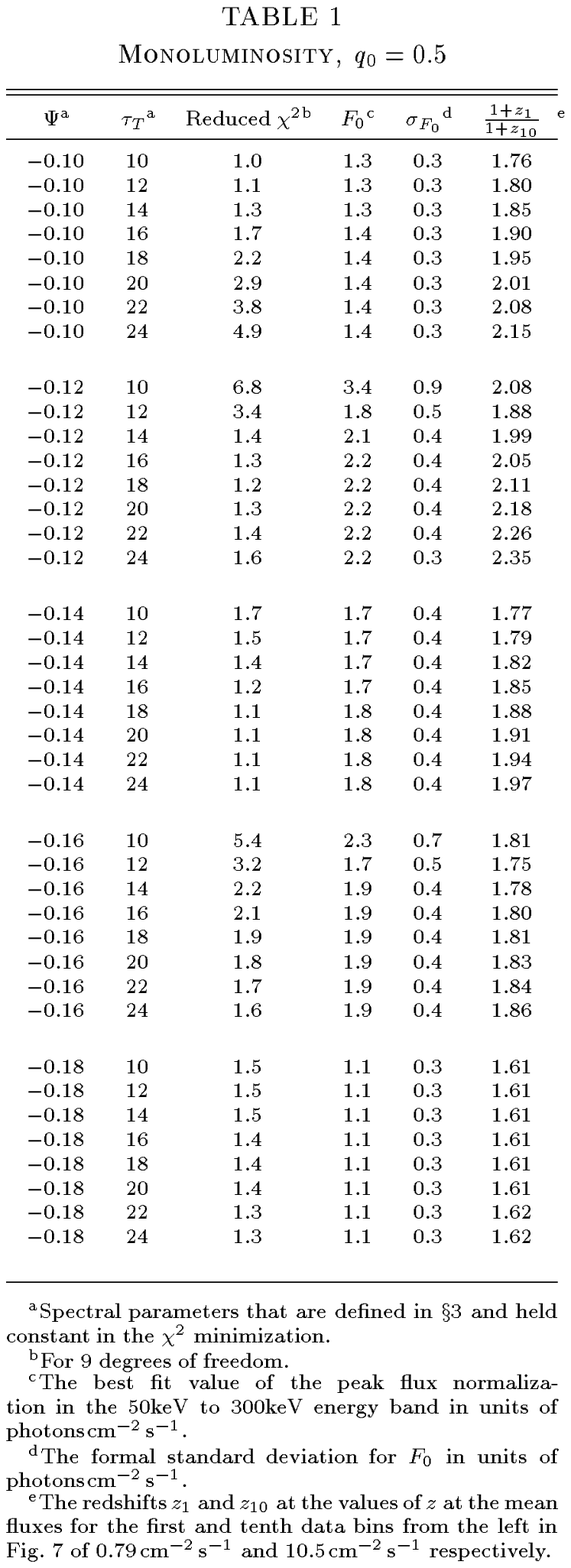}\hfill}
\end{figure}

\begin{figure}[tbp]
\hbox to \hsize{\hfill\epsfbox{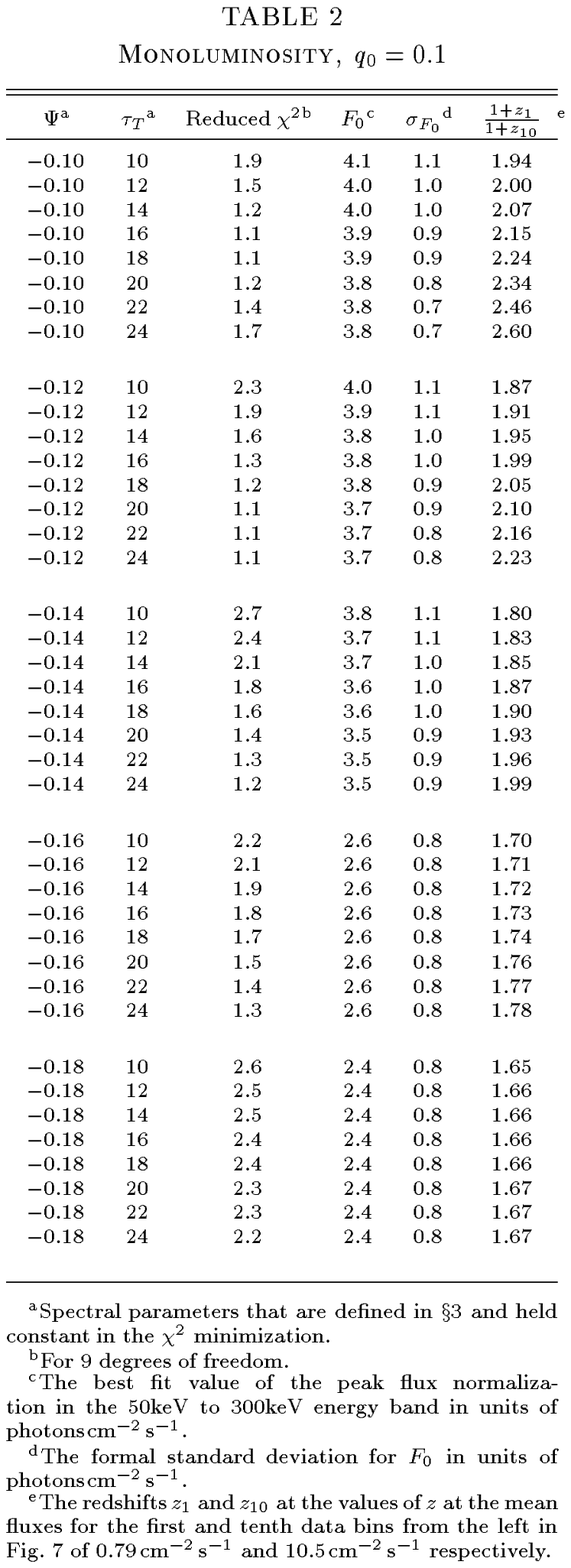}\hfill}
\end{figure}

\placefigure{fig7}

\begin{figure}[tbp]
\hbox to \hsize{
\epsfxsize=3.0in \epsfbox{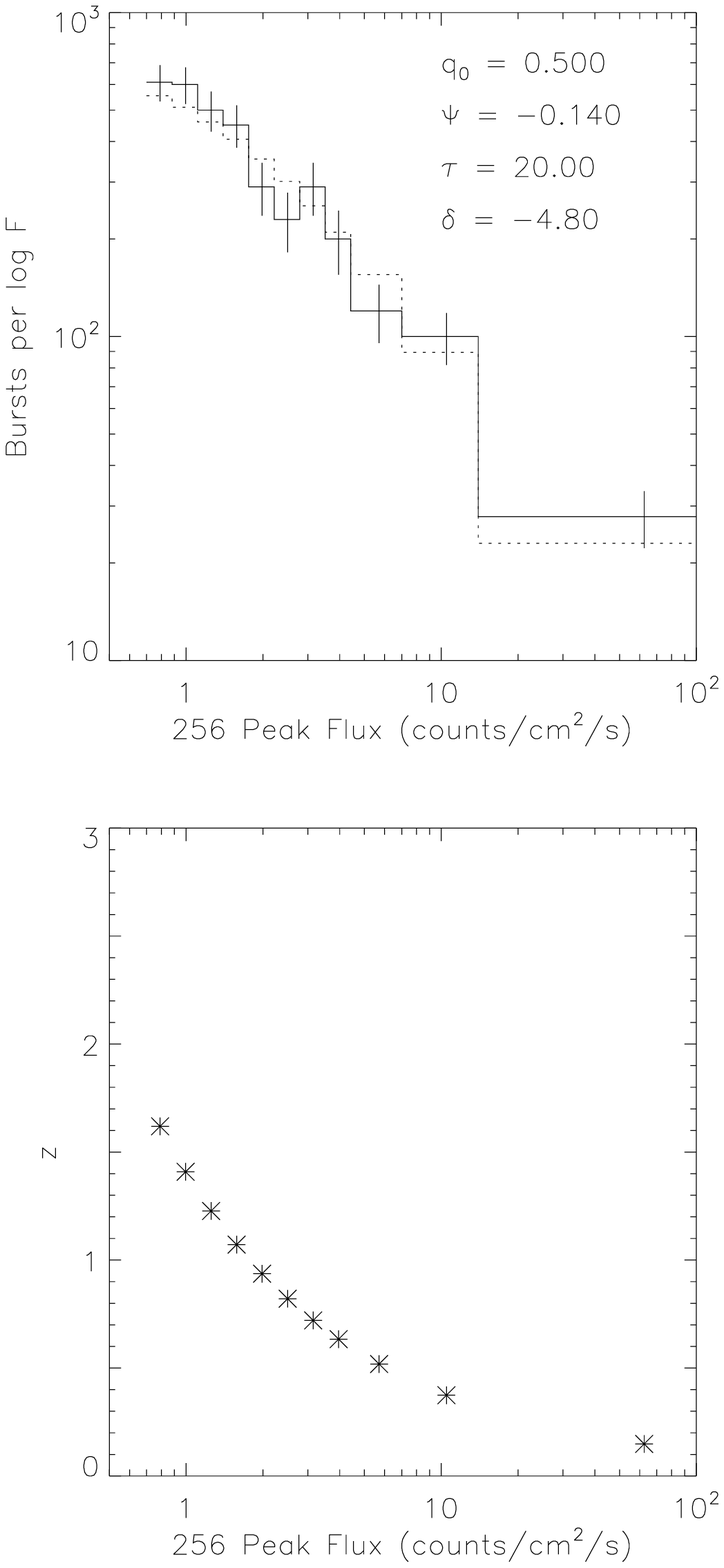}
\hfill
\epsfxsize=3.0in \epsfbox{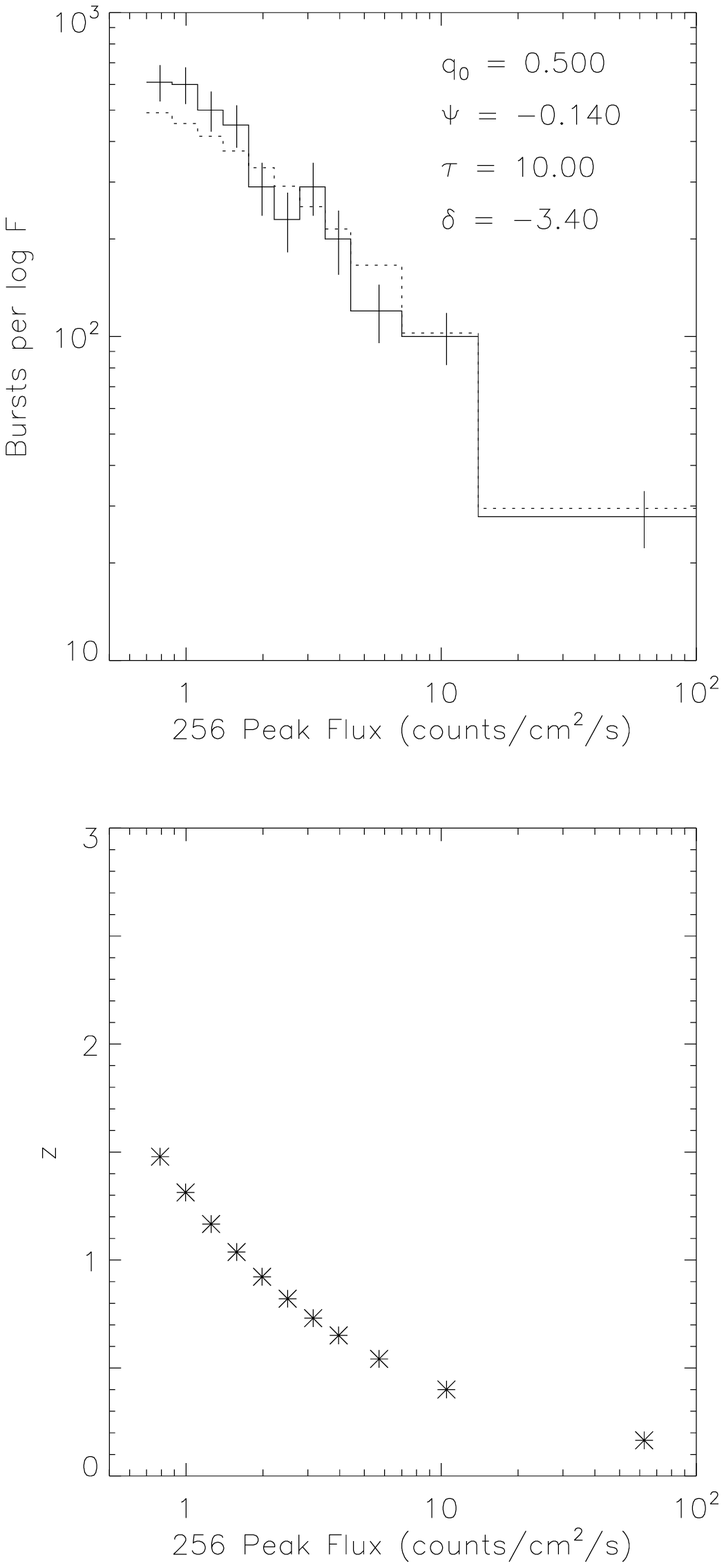}
}
\vskip 0.1in
\hbox to 5.5in { \hskip 0.5in
\begin{minipage}[b]{5.5in}
{\bf Fig. 7}---Examples of monoluminous peak-flux distributions
fit to BATSE data.  In the upper figures, the solid curve gives the BATSE
data with statistical errors and the dotted curve gives the expected value
from the best fit model.  The peak flux is for the $256 \, \ms$ timescale
in the $50 \, \keV$ to $300 \, \keV$ energy band.  The lower figures give
the value of $z$ as a function of peak flux for the best fit model.
For both models, $\Psi = -0.14$ and $q_0 = 0.5$.
On the left, the reduced $\chi^2$ is 1.1 for 9 degrees of freedom
with $\tau_T = 20$,
and, on the right, the reduced $\chi^2$ is~1.7 with $\tau_T = 10$.
\end{minipage}
\hfill
}
\end{figure}

In fits of the monoluminous flux distribution to the BATSE data,
I fix the values of $q_0$, $\Psi$, and $\tau_T$.
These variables are set by data that is independent of the peak
flux distribution. In particular, the values of $\Psi$ and $\tau_T$
are set by fitting gamma-ray burst spectra.  Generally one finds
$-0.16 < \Psi < -0.12$ and $10 < \tau_T < 30$ from fits to the spectra
of bursts observed by BATSE (\markcite{Brainerd5}Brainerd \etal\ 1996b).
This leaves two free parameters in the $\chi^2$ minimization, $F_0$
and $N_0$, giving 9 degrees of freedom in the minimization.
The results are given in Tables 1 and 2 for $\Psi$ ranging
from $-0.10$ to $-0.18$ and $\tau_T$ ranging from 10 to~26.
In these tables, the first and second columns give the values
at which $\Psi$ and $\tau_T$ are held, the third column gives
the value of reduced $\chi^2$ from the minimization,
the fourth column gives the value of $F_0$ derived from the 
minimization, and the fifth column gives the formal error on $F_0$.
The final column  gives the ratio of $1 + z_1$ to $1 + z_{10}$,
where, counting from the lowest to the highest peak-flux rate,
$z_1$ is the redshift at the center of the first data bin and $z_{10}$
is the redshift at the center of the tenth data bin.  These bins
are centered on the peak-flux rates of $0.79 \, \cm^{-2}\,\s^{-1}$
and $10.5 \, \cm^{-2}\,\s^{-1}$ respectively.
The minimum values of reduced $\chi^2$
given in the tables are plotted against $\tau_T$ in Figure~6.

The minimum value of $\chi^2$ is strongly dependent on the values chosen
for $\Psi$ and $\tau_T$, and only for specific combinations of
these values are good fits found.  The best value of reduced
$\chi^2$ is $\approx 1.0$, which has an expectation of $0.5$
for 9 degrees of freedom; an expectation of $\approx 0.01$ is found for
a reduced $\chi^2$ of 2.4.  For a given value of $q_0$, the value of
$\tau_T$ that gives a good fit increases rapidly as $\Psi$ decrease, that
is, as the value of $E_p$ decreases.
The characteristic values of $\tau_T$ and $\Psi$ one derives from
fits to the spectra ($\tau_T \approx 20$ and
$\Psi \approx -0.14$) produce a model fit to the flux distribution
with a reduced $\chi^2$ of 1.1 when $q_0 = 0.5$, and a
reduced $\chi^2$ of 1.9 when  $q_0 = 0.1$.
The values of $F_0$ that one derives vary little as $\Psi$
and $\tau_T$ vary;
it is 1--$2 \, \cm^{-2} \, \s^{-1}$ for $q_0 = 0.5$, and it is
2.5--$4 \, \cm^{-2} \, \s^{-1}$ for $q_0 = 0.1$.
As a consequence, the value of $z$ at the BATSE
threshold is $\approx 1.5$--$2.5$.

Figure~7 shows two specific examples of model fits, one of which fits
the data well, the other poorly.  When the model fits the data poorly,
it is because the model breaks more rapidly away from a $-5/2$ power law
than does the data.  These figures also show the variation of $z$ with~$F$.

\subsection{Cutoff Power-Law Luminosity Distribution}

\placefigure{fig8}
\placefigure{fig9}
\begin{figure}[tbp]
\hbox to \hsize{\epsfxsize=4.0in \epsfbox{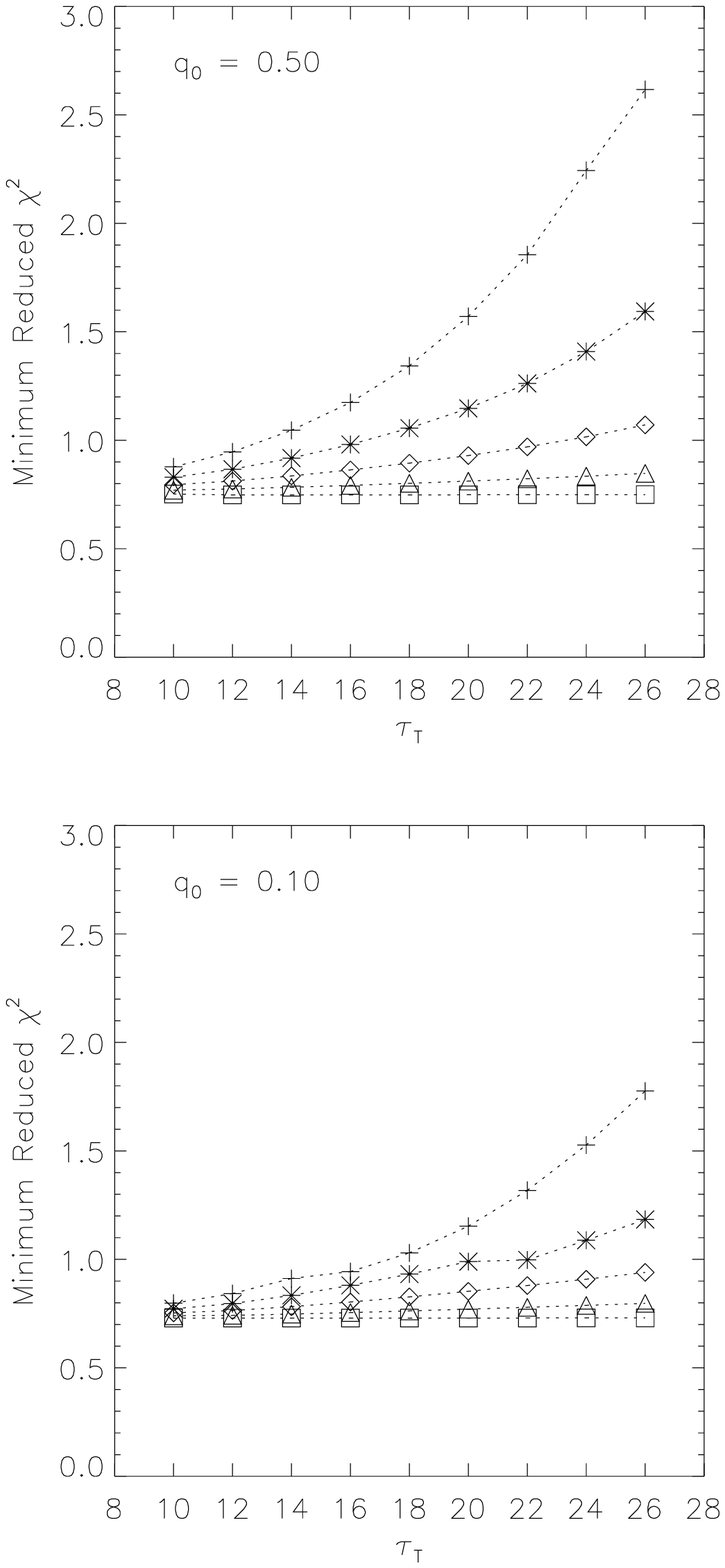}
\hskip 0.25in
\vbox to 9.0in {
\vfill
\begin{minipage}[b]{2.25in}
{\bf Fig. 8}---Reduced $\chi^2$ for fits of power-law luminosity
peak-flux distributions to BATSE data.  As in Fig.~6.
\end{minipage}
\vfill
}}
\end{figure}

\begin{figure}[tbp]
\hbox to \hsize{\hfill\epsfbox{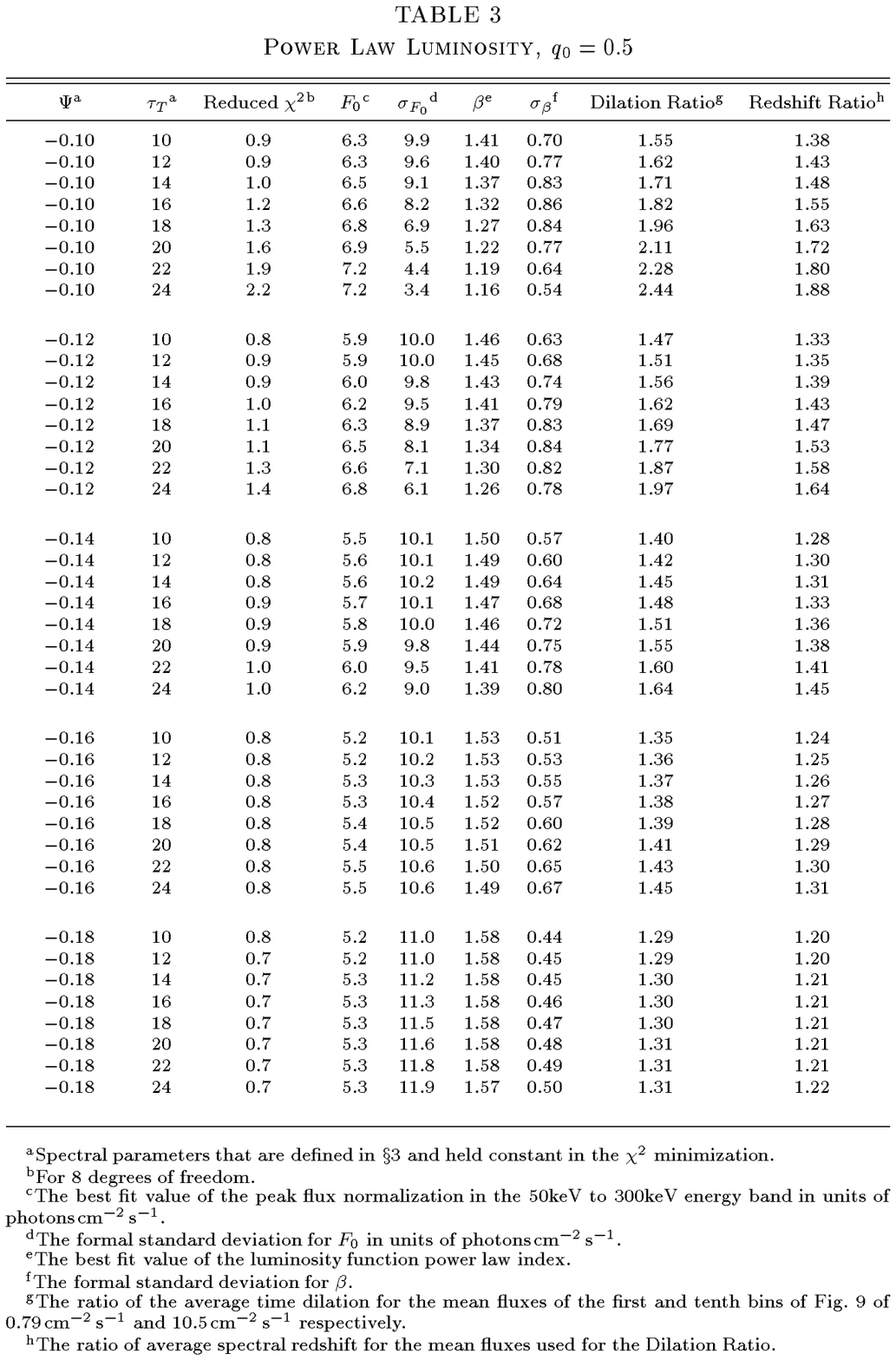}\hfill}
\end{figure}

\begin{figure}[tbp]
\hbox to \hsize{\hfill\epsfbox{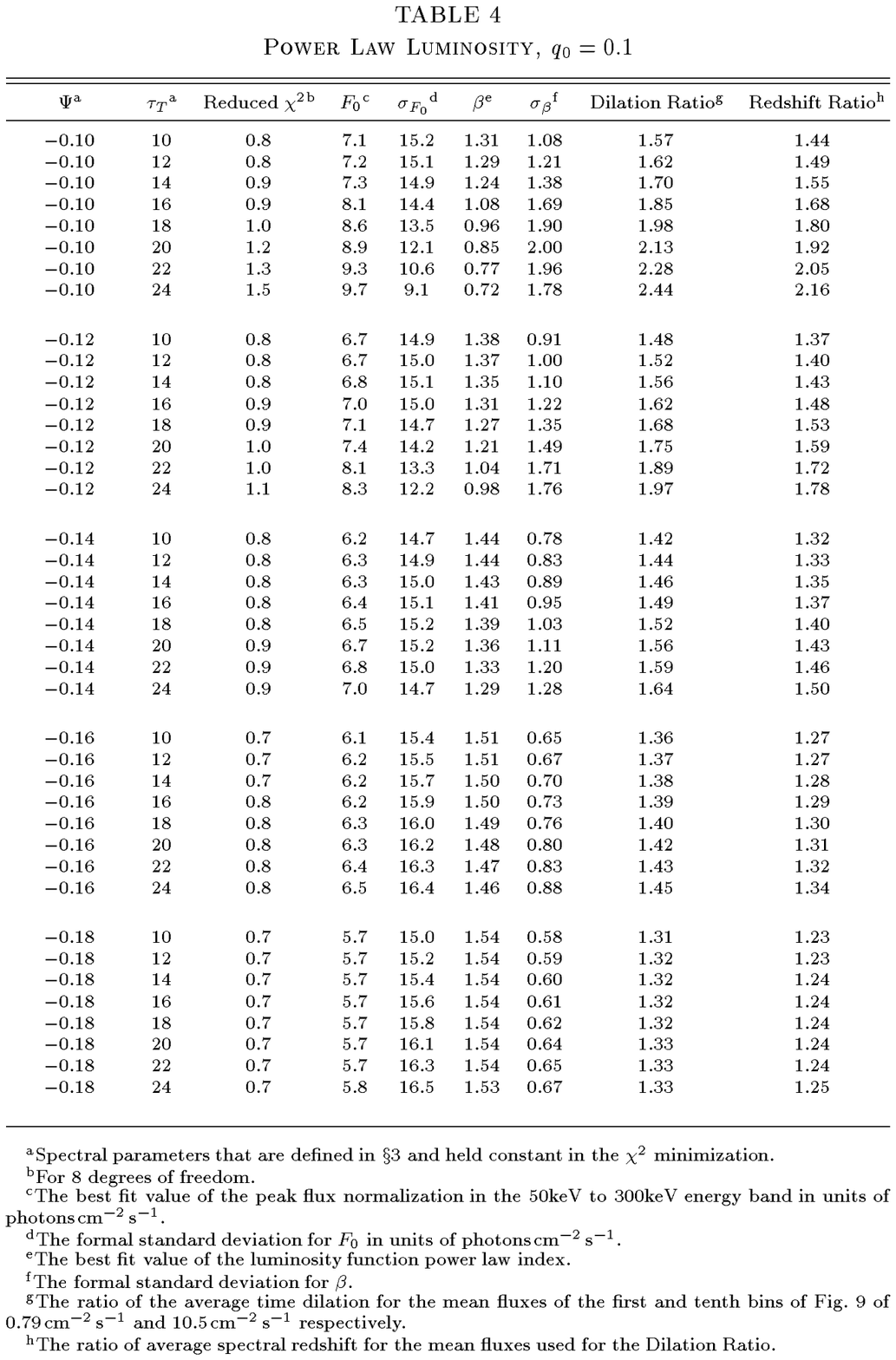}\hfill}
\end{figure}

\begin{figure}[tbp]
\hbox to \hsize{
\epsfxsize=3.0in \epsfbox{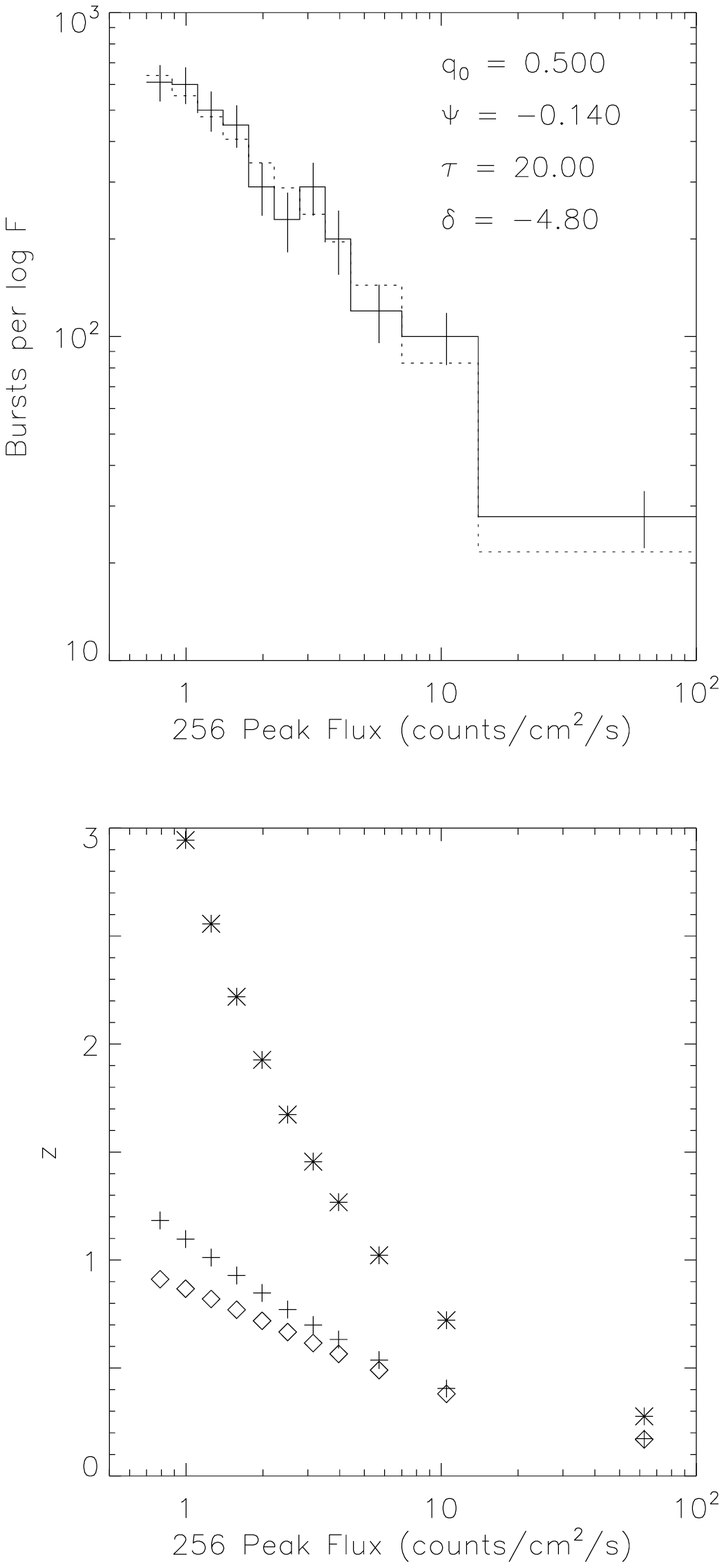}
\hfill
\epsfxsize=3.0in \epsfbox{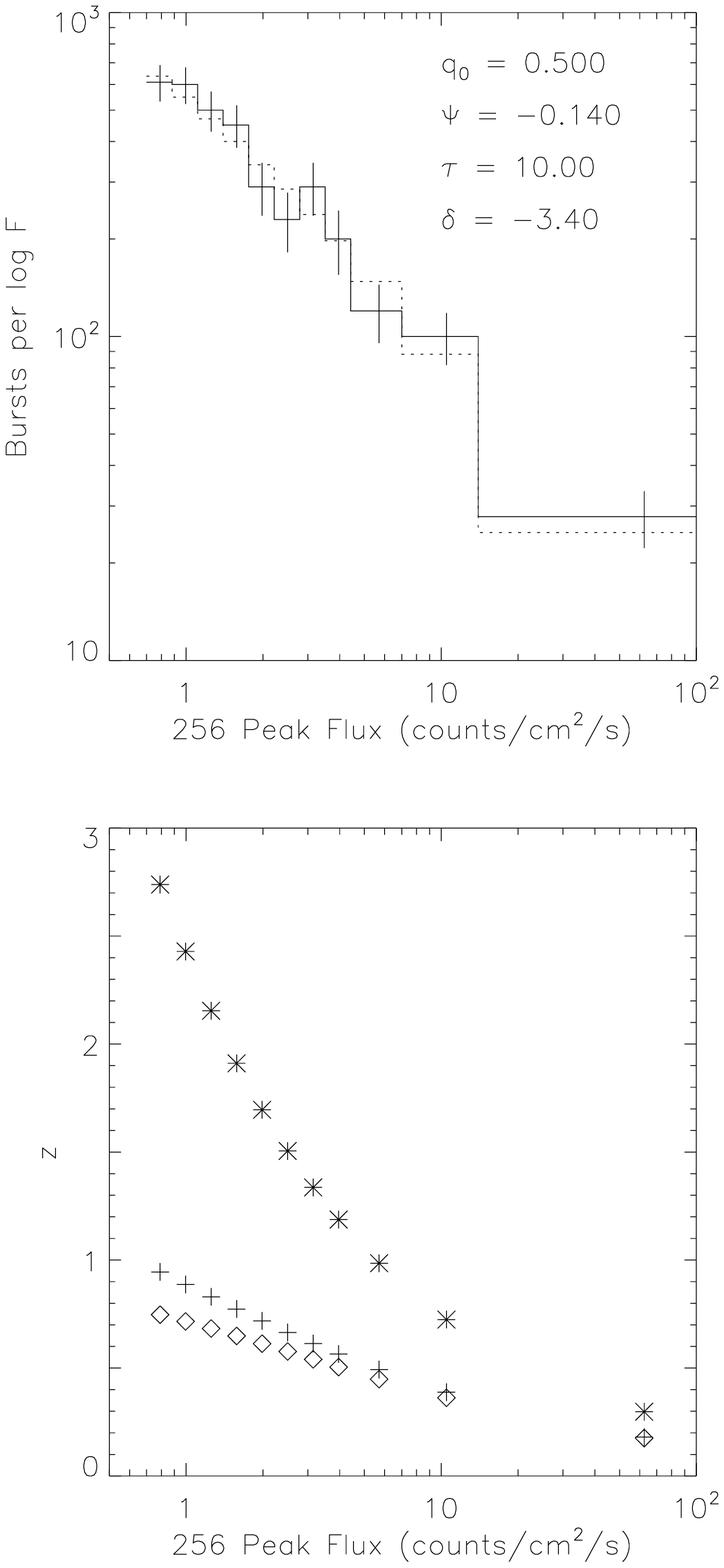}
}
\vskip 0.1in
\hbox to 5.5in { \hskip 0.5in
\begin{minipage}[b]{5.5in}
{\bf Fig. 9}---Examples of power-law luminosity peak-flux
distributions fit to BATSE data.  Upper figures are as in Fig.~7.
In the lower figures, the value of $z_{max}$ is given by $\star$,
the value of $\left< z \right>$ is given by $+$, and the value of
$1/\left< \left( 1 + z \right)^{-1} \right> - 1$ is given by $\Diamond$.
On the left, the reduced $\chi^2$ is 0.9 for 8 degrees of freedom with
$\tau_T = 20$, and on the right, it is 0.8 with $\tau_T = 10$.
The value of $z_{min}$ for the lowest bin in the
lower left figure is 3.39, which is off the plot.
\end{minipage}
\hfill
}
\end{figure}

The flux distribution for the power-law luminosity distribution
has three free parameters: $F_0$, $N_0$, and $\beta$.  Because
$q_0$, $\Psi$, and $\tau_T$ are fixed parameters, the model
fits have 8 degrees of freedom.  The introduction of a power-law luminosity
distribution improves the ability of a cosmological gamma-ray burst model
to fit the observed flux distribution, with most values of
$\Psi$, $\tau_T$, and $q_0$ producing acceptable values of
reduced $\chi^2$.  These values are plotted in Figure~8,
and the results of the model fits are listed in Tables 3 and~4.
The first five columns in these tables are as in Tables 1 and~2.
The sixth column is the derived value of $\beta$, the seventh
column is the formal error for $\beta$, the eight column is the ratio
of the average time dilation factor for the first data bin to the average
time dilation factor for the tenth data bin, and the tenth column
is the ratio of the average redshift factor for the first data bin to the
average redshift factor for the tenth data bin.
Two of the model fits listed in Table~3 are plotted in Figure~9.
For the values of $\Psi$ and $\tau_T$ consistent with observed
gamma-ray burst spectra, fits to the flux distribution produce
values of reduced $\chi^2 \approx 0.8$.

The derived values of $F_0$ range between $5 \, \cm^{-2} \, \s^{-1}$ and
$10 \, \cm^{-2} \, \s^{-1}$, with the best values of $\chi^2$ having the
lowest values of $F_0$.  But the formal errors for $F_0$ are of
order $10 \, \cm^{-2} \, \s^{-1}$.  Such values for
$F_0$ imply isotropic burst
luminosities of $\approx 2 \times 10^{51} \, \ergs \, \s^{-1}$
when $H_0 = 75 \, \km \, \s^{-1} \, \Mpc^{-1}$ in the $50 \, \keV$
to $300 \, \keV$ energy band.  The value of
$\beta$ ranges between $0.7$ and 1.6, with the highest values occurring
for the lowest values of $\chi^2$, but again the error is large, of order
0.5.  These values of $\beta$ are somewhat smaller than the value
of $\approx 1.9$ one would get from a pure power-law fit to the data.
The data therefore do not go to sufficiently low flux to exhibit
the asymptotic limit. The values of $\beta$ found in this study imply
that the number of bursts per decade luminosity changes slowly with
luminosity.  For instance, with $\beta = 1.6$, only 75\% of the bursts
are within a factor of 10 of the lower threshold.
This shows that gamma-ray bursts do not need a steeply falling
or a narrow luminosity distribution to reproduce the observations.

The large values of $\beta$ in these fits imply that the redshift
distribution function is strongly peaked at $z < 1$.  The effect of
this is clear in the lower halves of Figures~9a and 9b.  The value of
$z_{max}$ for the bursts at threshold is generally $\gtrsim 3$.  On the
other hand, the average time dilation and redshift factors are relatively
small and slowly varying.  In Figure~9, the implied redshifts from
averages over $\left( 1 + z \right)$ and $1/\left( 1 + z \right)$ are
plotted.  Even though $z_{max} \gtrsim 3$, the average redshifts are
$\lesssim 1$.  Comparing these figures to Figure~7 shows that the variation
of the average $z$ with $F$ for the power-law luminosity model is less
than the variation of $z$ for the monoluminous model.  For instance,
in Figure~7a, the lowest flux bin, which is centered at
$0.79 \, \cm^{-2}\, \s^{-1}$, has $z = 1.62$ and the tenth bin, which
is centered at $10.5 \, \cm^{-2}\, \s^{-1}$, has $z = 0.37$; in contrast,
in Figure~9a, the average values of $z$ from the time dilation are
$1.18$ and $0.41$ for these flux bins, and the average values of
$z$ from the spectral redshift are $0.91$ and $0.38$.

The values of $z$ derived from fits to the peak flux distribution
are lower but consistent with the values of $z$ derived from fits
of the Compton attenuation spectral model to the spectra of bursts
observed by BATSE.  Spectral fits to relatively bright bursts (peak fluxes
$> 10 \, \cm^{-2} \, \s^{-1}$) produce $0.5< z < 4$
(\markcite{Brainerd5}Brainerd \etal\ 1996b).
Given the large uncertainty in $F_0$, those results do not contradict
the results presented above.

Three conclusions are evident.  First, fits of the data to the monoluminous
model are more sensitive to the spectral parameters
than are fits to the power-law luminosity models.  Second, the maximum
redshifts that are consistent with the data are much larger when a
luminosity function is present than when bursts are monoluminous.
Third, the spectral redshift and time dilation from the cosmological
expansion are weakened by the presence of a luminosity distribution.


\section{Redshift and Time Dilation Tests}

\placefigure{fig10}
\begin{figure}[tbp]
\hbox to \hsize{
\epsfxsize=3.0in \epsfbox{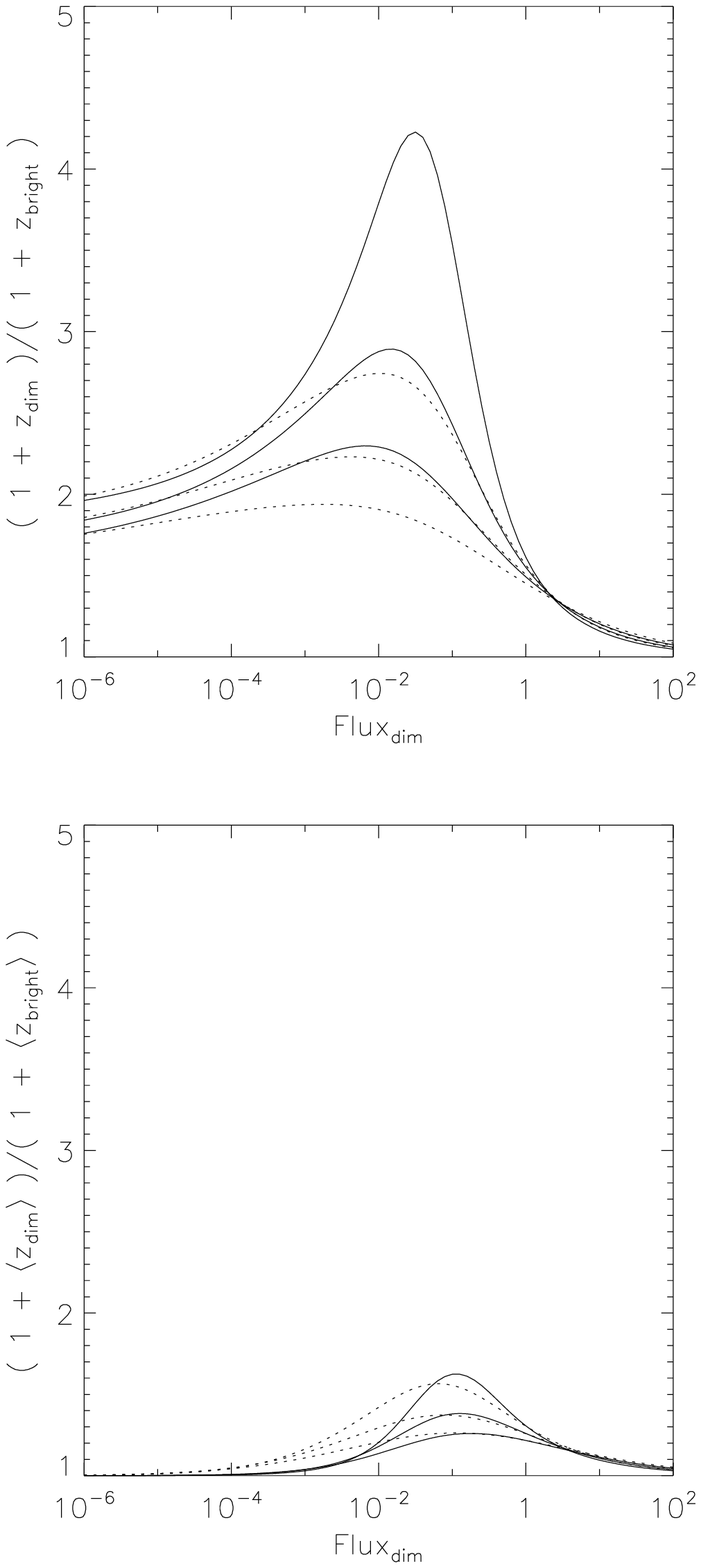}
\hfill
\epsfxsize=3.0in \epsfbox{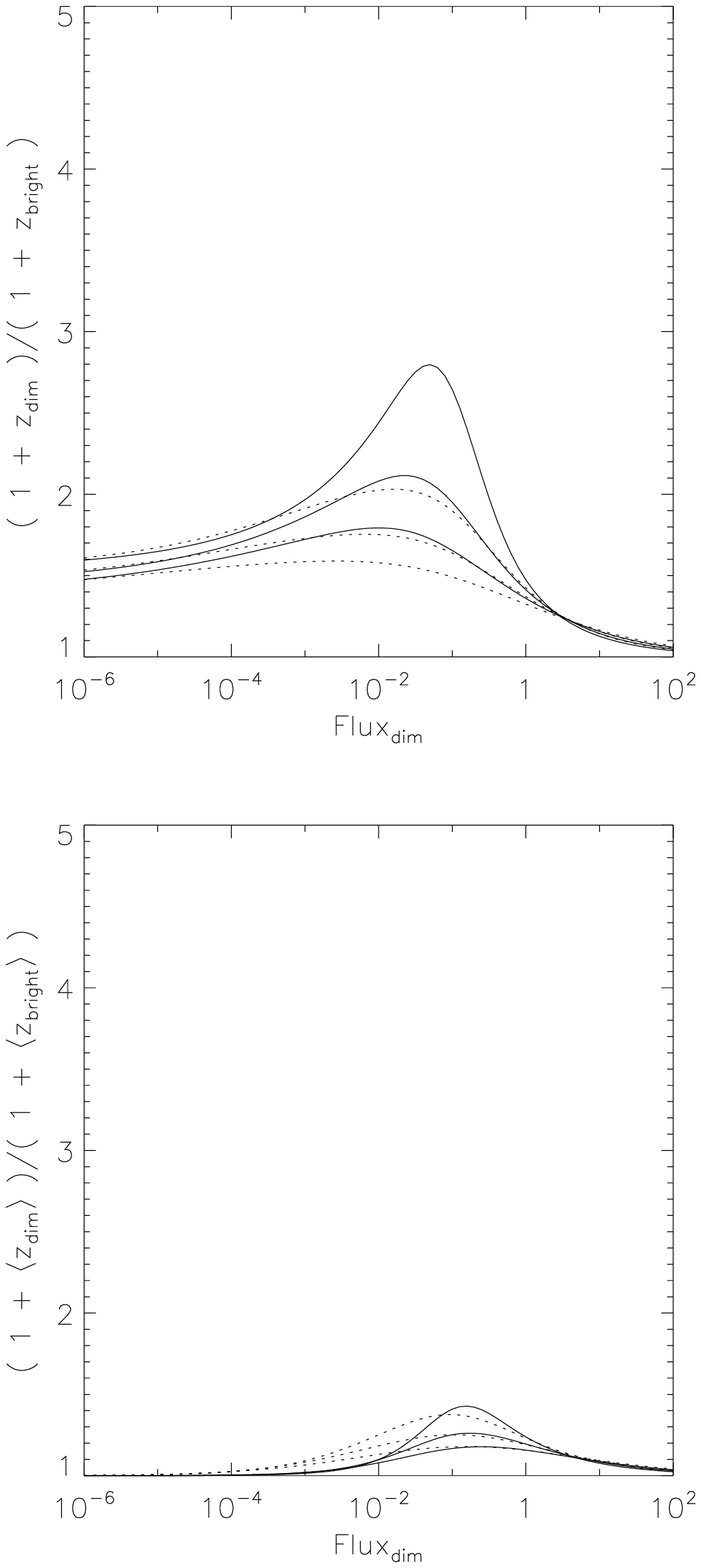}
}
\vskip 0.1in
\hbox to 5.5in { \hskip 0.5in
\begin{minipage}[b]{5.5in}
{\bf Fig. 10}---$\left( 1 + z_{dim} \right)/\left( 1 + z_{bright} \right)$
versus peak flux.  The ratios of time dilation factors for various
cosmological models are plotted.  The cosmological redshifts $z_{dim}$
and $z_{bright}$ are the redshifts associated with $F_{dim}$ and
$F_{bright}$, where, on the left, $F_{bright} = 10 \, F_{dim}$ and,
on the right, $F_{bright} = 5 \, F_{dim}$.  The upper figures are
for the monoluminous model, with the solid lines for $q_0 = 0.5$ and
the dotted lines for $q_0 = 0.1$.  The lower figures are for the
power-law luminosity model with $\beta = 1.6$ and the remaining
parameters as in the upper figures.  For these figures,
$z_{dim}$ and $z_{bright}$ are average values of redshift. 
 For all curves $\tau_T = 18$, and, from top to bottom in each figure,
the curves have $\Psi = -0.1$, $-0.14$, and $-0.18$.  The flux
is normalized so that $F_0 = 1$, and it is measured in the $50 \, \keV$
to $300 \, \keV$ energy band.
\end{minipage}
\hfill
}
\end{figure}

Researchers have sought evidence of a cosmological time dilation
and spectral redshift in sets of gamma-ray bursts by comparing the
characteristics of bursts at one flux to those at another.  Such
studies derive the ratio of the average time dilation (or redshift)
at these two fluxes rather than the average value of~$z$
(\markcite{Mitrofanov2}Mitrofanov \etal\ 1994,
\markcite{Mitrofanov3}1996; \markcite{Fenimore2}Fenimore \& Bloom 1995).
When the bursts are monoluminous, this ratio goes to a constant
as $F \rightarrow \infty$.  From equation (7), with the spectrum set
to a power law $\propto \nu^{\delta}$, the limiting value of the redshift
ratio $\left( 1 + z_{dim} \right)/\left( 1 + z_{bright} \right)$
for the fluxes $F_{dim} < F_{bright}$ is
$\left( F_{bright}/F_{dim} \right)^{1/\left( 2 - \delta \right)}$
when $1 \ll z \ll 1/q_0$ and $\left( F_{bright}/F_{dim} \right)^{-1/\delta}$
when $z \gg 1/q_0$.  For a flux ratio of $F_{bright}/F_{dim} = 10$ and a characteristic power-law index of $\delta = -3$, the first relationship
gives $2.15$ and the second gives $1.58$.   The upper plot of Figure~10a
gives the behavior of
$\left( 1 + z_{dim} \right)/\left( 1 + z_{bright} \right)$
as a function of $F_{dim}$ for $F_{bright}/F_{dim} = 10$.
The asymptotic behavior for small flux
is evident in this figure.  More striking is the maximum in each curve
at $F \approx 3 \times 10^{-2} F_0$.
The maximum is generally $\lesssim 4$, with the
value of the peak decreasing with $\Psi$ and $q_0$.  The values of this
ratio for the model fits discussed above are given in Tables 1 and 2 for
$F_{dim} = F_1 = 0.79 \, \cm^{-2} \, \s^{-1}$, the flux at the center of
the first flux bin, and $F_{bright} = F_{10} = 10.5 \, \cm^{-2} \, \s^{-1}$,
the flux at the center of the tenth flux bin.
The value of $z$ at the center of bin 1 is $z_1$, and its value at the
center of bin 10 is~$z_{10}$. Note that this ratio
is $\lesssim 2$ in the tables.  Figure 10b gives the results for
$F_{bright}/F_{dim} = 5$.  Lowering the ratio of fluxes dramatically
decreases the redshift ratio curves.

Introducing a power-law luminosity function lowers the value of the average
time dilation and redshift ratios from those found in the monoluminous model.
In the lower plot of Figure~10a, one sees that the ratios of average
time dilation are much smaller than in the monoluminosity case and that
the peak value of the ratio is generally $\lesssim 1.7$.  The time dilation
and redshift ratios for the first and tenth bins for the model fits to
the data are given in Tables~3 and~4.  The time dilation and redshift ratios
for values of reduced $\chi^2$ near 0.8 are $\lesssim 1.6$.  These values
can be as low as $\approx 1.20$ for some spectral parameters that produced
good model fits to the flux distribution.  This demonstrates that the
cosmological effects are never large when a luminosity distribution
determines the shape of the flux distribution, and that often the
cosmological effects are slight, even though many bursts in the model
have $z \gtrsim 1$.  Again, going to a flux ratio of $F_{bright}/F_{dim} = 5$
dramatically lowers these ratio curves.

Many studies report a correlation of the gamma-ray burst timescale
with the peak flux, finding that the dim bursts are on average
longer than the bright bursts.  Examples of the tests producing
positive results are wavelet analysis, the total normalized
counts test (\markcite{Norris1}Norris \etal\ 1994),
and the fitted pulse width test
(\markcite{Davis}Davis \etal\ 1994).  One test, the peak aligned test
(\markcite{Mitrofanov1}Mitrofanov \etal\ 1993) produces conflicting
results, with some researchers finding a correlation in the BATSE data
(\markcite{Norris1}Norris \etal\ 1994), and others finding no correlation
in either the BATSE data (\markcite{Mitrofanov3}Mitrofanov \etal\ 1996)
or the data from the APEX experiment on Phobos-2
(\markcite{Mitrofanov1}Mitrofanov \etal\ 1993).  This is somewhat
disheartening, because clear evidence of a correlation would give
the discussion that follows more resonance, but there is no consensus
on the issue, so I'll address both sides of it.

Those timescale studies that report a correlation generally find
a dramatic effect.  Norris \etal\ \markcite{Norris1} (1994) find
that BATSE gamma-ray bursts with peak count rates above
$18000 \, \s^{-1}$ are $\approx 2.25$ times as long as bursts
with peak count rates between $1400 \, \s^{-1}$ and $2400 \, \s^{-1}$.
The actual time dilation ratio required to produce such an effect
is much larger, because gamma-ray burst durations are photon-energy
dependent.  If the duration is proportional to $\nu^{-0.23}$
(\markcite{Mitrofanov3}Mitrofanov \etal\ 1996), then a time dilation
ratio of $2.8$ is required.  From Figure 10, one sees that a ratio of
this size is impossible to achieve in the power-law luminosity model,
and it is possible in the monoluminous model for only part of the
available parameter space.  As \markcite{Fenimore2}Fenimore \& Bloom (1995)
note, the monoluminous models produce a time dilation ratio of the
required size at $z_{dim} \gtrsim 6$, which is inconsistent with the
flux distribution.  When the time dilation ratios for a factor of 10
difference in flux are calculated for the models in Tables 1--4, one
finds that the monoluminous model produces ratios between 1.5 and~2
and the power-law luminosity model produces ratios between 1.2
and~1.5.

To be more precise about this, let us use the specific model of
a $q_0 = 0.1$ universe and a $\Psi = -0.14$ and $\tau_T = 20$ spectral
model.  For a power law luminosity, a fit to the peak flux distribution
gives $F_0 = 6.7 \, \cm^{-2} \, \s^{-1}$ and $\beta = 1.36$, while for
a monoluminous distribution, $F_0 = 3.5 \, \cm^{-2} \, \s^{-1}$.

In Davis \etal\ (\markcite{Davis}1994), time dilation ratios of $1.9^{+0.34}_{-0.44}$, $1.8^{+0.38}_{-0.33}$, and $1.6^{+0.39}_{-0.47}$
are found from three different statistical tests when comparing bursts
with count rates between 18000 and 250000 counts per second to those
between 2400 and 4500 counts per second.  These tests are all based on
fitting with width of pulse structure within a burst, and are therefore
dependent on the photon energy.  Taking this lower count
rate to be a photon flux of $0.45 \, \cm^2 \, \s^{-1}$, and taking
the lower edge of the upper interval to be $5.8 \, \cm^2 \, \s^{-1}$,
and assuming that the width of the pulses go as $\nu^{-0.23}$, one finds
dilation ratios of 1.52 for the monoluminous model and 1.31 for the
power-law luminosity model.  Comparing the same upper count band to
a lower range of 1400 to 2400 counts per second, the tests give ratios
of $2.0^{+0.62}_{-0.47}$, $2.2^{+0.72}_{-0.44}$, and $1.8^{+0.65}_{-0.51}$,
while the monoluminous model gives 1.52, and the power-law luminosity
model gives 1.41. These various tests produces results that are all
larger than the monoluminous model by about one standard deviation,
and larger than the power-law luminosity model by about one and one-half
standard deviations.  For the quoted errors, the monoluminous model
produces a value that is one and one-half standard deviations above
unity, and the power-law luminosity model produces a value that is
less than one standard deviation from unity.

A test that compares the time between peaks, which is not energy
dependent to the extent that the width of a single peak is, finds
dilation ratios of 2.18 and 2.20 on the $256 \, \ms$ timescale for
two different selection criterion, with significances compared to
unity of 0.013 and 0.0016 respectively
(\markcite{Norris2}Norris \etal\ 1996).
Assuming gaussian statistics, these significances correspond to 2.5
and 3.2 standard deviations from unity, so the standard deviations
for these measures are 0.47 and 0.38 respectively.  In the absence
of energy dependence of the measured duration, the monoluminous model
gives 1.73 and the power-law luminosity model gives 1.42. The observed
result is of order a standard deviation above the monoluminous model and
two standard deviations above the power-law luminosity model.

The one test that finds no evidence of a duration--peak flux correlation
is the analysis of Mitrofanov \etal\ \markcite{Mitrofanov3} (1996).
The peak-aligned method is used in this article to compare the average
duration of bursts with different peak fluxes.  It tests a different
aspect of burst timescale than those just discussed, so the absence of
a signal does not necessarily contradict these other tests.  There
is a contradiction with Norris \etal\ (1994), which Mitrofanov \etal\ show
to be a consequence of the selection criterion use by Norris \etal.
Mitrofanov \etal\ report on several measures of burst duration.
In the first, bursts with peak fluxes $< 1 \, \cm^{-2} \, \s^{-1}$
are compared to bursts with peak flux $> 1 \, \cm^{-2} \, \s^{-1}$.
The total widths of the averaged light profiles are $6.57 \pm 0.09 \, \s$
and $6.64 \pm 0.1 \, \s$ respectively.  This implies a duration ratio
of $0.99 \pm 0.13$.  In comparison, one finds a time-dilation ratio
of $1.22$ for a monoluminous model with a width energy dependence of
$\nu^{-0.23}$, and a time-dilation ratio of $1.12$ for a power-law
luminosity model with the same energy dependence.  The data is two
standard deviations from the first, but it is only one standard
deviation from the second.  In the second, bursts with peak fluxes
$< 0.45 \, \cm^{-2} \, \s^{-1}$ are compared to bursts with peak
flux $> 2.5 \, \cm^{-2} \, \s^{-1}$.  These bursts give total widths
of $6.50 \pm 0.13 \, \s$ and $6.21 \pm 0.15 \, \s$ respectively.
The duration ratio is then $1.05 \pm 0.20$.  Taking
$0.45 \, \cm^{-2} \, \s^{-1}$ for the low flux, one finds
$1.49$ for the monoluminous models and $1.27$ for the power-law
luminosity model.  This data is a little over two standard deviations
from the first model value and one standard deviation from the second
model value.  From this, one sees that the data can be considered
consistent with a cosmological time dilation if bursts have a
broad luminosity distribution function.

One study attempted to find a correlation between the peak flux 
and $E_{peak}$, the observed photon energy at which the $\nu F_{\nu}$
curve has a maximum (\markcite{Mallozzi1}Mallozzi \etal\ 1995).
The advantages of using $E_{peak}$ rather than burst timescale
are that $E_{peak}$ is strictly proportional to $1/\left( 1 + z \right)$,
and its value varies by less than a factor of 10, whereas the duration
is energy dependent, and it varies in value by several orders of magnitude.
The study found that the average value of $E_{peak}$ increases with
peak flux, which one expects from cosmological effects.  But the
implied redshift factor is large. The subset of BATSE gamma-ray bursts
with T90 $> 2 \, \s$ produces a ratio of average $E_{peak}$ of
$2.45^{+0.25}_{-0.66}$ for bursts with peak fluxes between
$5.90 \, \cm^{-2} \, \s^{-1}$ and $105.0 \, \cm^{-2} \, \s^{-1}$
relative to bursts with peak fluxes between
$0.95 \, \cm^{-2} \, \s^{-1}$ and $1.30 \, \cm^{-2} \, \s^{-1}$.
In contrast, model calculations for the parameters given above produce
1.61 for the monoluminous model and 1.30 for the power law luminosity
model, which are 1.3 and 1.7 standard deviations from the data.
For a lower peak flux range of $0.95 \, \cm^{-2} \, \s^{-1}$ to
$1.30 \, \cm^{-2} \, \s^{-1}$, the ratio is $1.99^{+0.36}_{-0.38}$,
while the models give 1.47 and 1.23 for the monoluminous
and the power law luminosity cases, which are 1.4 and 2.0 standard
deviations from the data.

Taken individually, the tests that show a correlation of burst
duration or characteristic spectral energy with peak flux are
consistent with the models, but taken together, they are inconsistent,
because all of these tests are systematically above the monoluminous
model by one standard deviation and above the power-law luminosity
model by two standard deviations.  The reported correlations are
therefore inconsistent with the theoretical models presented in this
article.  In conclusion, the reported correlations must be a
consequence of an intrinsic correlation of $E_{peak}$ and burst duration
with peak flux.  No cosmological model can simultaneously fit the
observed flux distribution and the claimed time dilation and redshift
ratios, and no cosmological model with a power-law luminosity distribution
can fit the claimed time dilation and redshift
ratios for any value of~$F_0$.


\section{The Flux Distribution for Redshift-Selected Subsets}

The observations require an intrinsic correlation of $E_{peak}$ and
time dilation with luminosity.  The question is whether a direct measure
of this correlation exists.  The recent studies of the flux distribution
of gamma-ray bursts selected by hardness ratio presents such an opportunity.

The discussion that follows depends on the two critical assumptions
underlying equation~(8).  First,
for $z \ll 1$, $n_0 \left( \tau \right)$ is a constant.  Second,  the
gamma-ray burst characteristics must be independent of $\tau$, particularly
for $z \ll 1$.  These assumptions are reasonable given that the
observed flux distribution goes to a power-law of index $-5/2$ for large flux,
which is a natural consequence of these assumptions.

The discussion in \S 3 and \S 4 shows that the power-law index of the flux
distribution deviates from $-5/2$ at $F_0$.  This flux is dependent on
the spectral parameters through ${\cal F}\left( L, {\vec p} \, \right)$
in the $L$ argument of $\Phi\left(L,{\vec p} \, \right)$, but this
dependence is weak, as demonstrated by the values for $F_0$ derived from
model fits and given in Tables 3 and~4.  If there is a strong
dependence of $F_0$ on spectral parameter, it can only enter through the
direct dependence of the distribution function
$\Phi\left(L, {\vec p}\,\right)$ on the spectral parameter ${\vec p}$.
In this case, $L_0$ is a function of ${\vec p}$, which makes $F_0$
a function of ${\vec p}$ through equation~(6).

\placefigure{fig11}
\begin{figure}[t]
\hbox to \hsize{\epsfxsize=4.0in \epsfbox{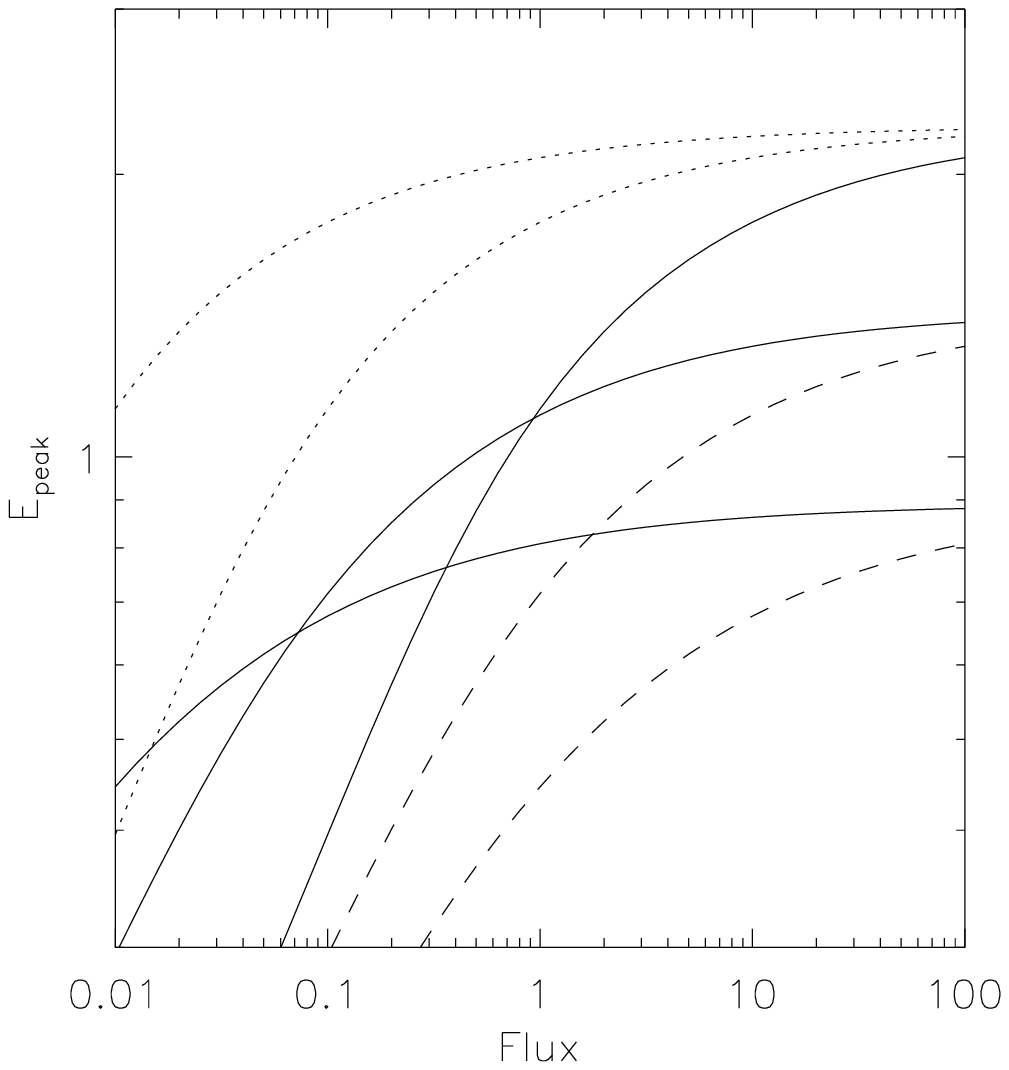}
\hskip 0.25in
\vbox to 4.0in {
\vfill
\begin{minipage}[b]{2.25in}
{\bf Fig. 11}---Models of $E_{peak}$ versus
$F$ for constant $L$.  From top to bottom, the solid lines are for
$\Psi = -0.12$ and $F_0 = 1$, $\Psi = -0.14$ and $F_0 = 0.1$,
and $\Psi = -0.16$ and $F_0 = 0.01$.  From top to bottom, the
dotted lines are for $F_0 = 0.01$ and $F_0 = 0.1$ with $\Psi = -0.12$.
From top to bottom, the dashed lines are for $\Psi = -0.14$ and $-0.16$
with $F_0 = 1$ and $F_0 = 0.1$.  The flux is measured in the $50 \, \keV$
to $300 \, \keV$ energy band.
\end{minipage}
\vfill
}}
\end{figure}

The dependence of $F_0$ on $L$ is graphically demonstrated in Figure~11,
where $E_{peak}$ is plotted as a function of flux for specific values
of $L$ and $E_p$, where $E_p$ is
the value of $E_{peak}$ when $z = 0$.  The solid
curves give the variation of $E_{peak}$ with $F$ when $E_p$ decreases
with $L$, the dotted curves give $E_{peak}$ for different values of $L$
and the same $E_p$, and the dashed curves give $E_{peak}$ for different
values of $E_p$ and the same $L$.
In these curves, $z \ll 1$ when $E_{peak}$ is
constant with $F$, so the flux distribution for a given $E_{peak}$ is
a power law of index $-5/2$.  Regarding these sets of curves as
three models of the dependence of $L_0$ on $E_p$, one
sees that when $E_p$ decreases with $L_0$,
the value of $F_0$ for bursts selected by  $E_{peak}$ decreases with
$E_p$.  The solid curves exhibit a second
property not yet touched upon---the crossing of curves. This means that
bursts with a given $F$ and $E_{peak}$ can have a variety of values for
$L$ and $z$.

For the spectral model in \S 2, the selection of bursts by
$E_{peak}$ is equivalent to turning the spectral
parameter $\Psi$ into a function of $E_p$ and $z$.
Equation (8) becomes
\begin{eqnarray}
	{d N \over d t \; d \Omega \; d F \; d z \; d E_{peak} \; d \tau_T }
   = n_0 \left[ \tau\left(z\right) \right] \, 
   &\Phi\left\{ { F \over {\cal F}\left( z \right) },
	\Psi\left[ E_{peak} \left( 1 + z \right), \tau_T \right] \right\}
   \, {\cal F}\left( z \right)^{-1} \nonumber \\
   &\cdot { r^2 \over \left( 1 + z \right) \, \sqrt{ 1 + 2 q_0 z } }
	\left| { d \Psi\left(E_p\right)
	\over d E_p } \right|_{E_{peak}\left( 1 + z \right)}
   \, .
\end{eqnarray}
The function $\Psi\left(E_p\right)$ is given by
$\Psi\left(E_p\right) = \sigma_c^{\prime}\left( E_p/m c^2 \right) E_p
	/\sigma_T m c^2$,
where $\sigma_c$ is the Klein-Nishina cross section and the prime
represents the first derivative with respect to $E_p/m c^2$.  For 
$E_p/m c^2 \gg 0.1972$, $\Psi\left(E_p\right) \approx
	\left[ \log\left( 2 E_p/m c^2 \right) - 1/2 \right]
	m c^2/E_p$,
so the first derivative of $\Psi\left(E_p\right)$ is
$d\Psi\left(E_p\right)/d E_p =
	\left[ \log\left( 2 E_p/m c^2 \right) - 3/2 \right] m c^2/E_p^2$.
Therefore, the derivative term does not dramatically
change the dependence of equation (22) on $z$ from that in equation~(8).
The important difference from equation (8) is the dependence of $\Phi$
on $E_{peak}\left( 1 + z \right)$ in equation (22).  The observed
distribution of $E_{peak}$ in the BATSE catalog falls rapidly to zero
for $300 \, \keV < E_{peak} < 1 \, \MeV$
(\markcite{Mallozzi1}Mallozzi \etal\ 1995).
Because the
$E_{peak}$ distribution is narrow, $\Phi\left( L, {\vec p} \, \right)$
must fall rapidly as $E_p$ goes to infinity.  This means that an
additional upper limit on the value of $z$ is set by the value of
$E_{peak}\left( 1 + z \right)$ at which the integrand no longer
contributes significantly to the integral.  When the integrand
peak occurs at $z_0 \ll 1$, the maximum value of $z$ is set by the
dependence of $\Phi$ on $L$, but when the peak occurs at $z_0 \gg 1$,
the maximum value of $z$ is set by the dependence on $E_p$.
In the second case, the upper limit on $z$ is independent of $F$.
This limit sets an upper limit on $L$ through the ratio
$F/{\cal F}\left( z, {\vec p}\,\right)$; as $F$ decreases, the upper
limit on $L$ decreases.

These results explain the results of recent studies
(\markcite{Pizzichini}Pizzichini 1994;
\markcite{Belli}Belli 1995;
\markcite{Kouveliotou}Kouveliotou \etal\ 1996;
\markcite{Pendleton2}Pendleton \etal\ 1997) of the BATSE flux distribution
for gamma-ray bursts selected by hardness ratio.
The hardness ratio is a measure of the spectral shape, and it is dependent
on the cosmological redshift because of the curvature in gamma-ray burst
spectra below $1 \, \MeV$.  It is therefore a proxy measure of $E_{peak}$.
These studies show that the soft gamma-ray bursts have a flux distribution
that is a power law of index $-5/2$.  The hard gamma-ray bursts, however,
follow a curve that falls below a $-5/2$ power law.
This is the behavior expected when $E_p$ rises with luminosity.

The study of the burst distribution on the $E_{peak}$--$\log F$ plane
may yield the luminosity distribution of gamma-ray bursts as a function
of $E_p$.  This would enable one to separate the intrinsic correlation
of $E_p$ with luminosity from the cosmological redshift.  The way one
approaches this problem is to select the group of burst with the highest
value of $E_{peak}$.  As one goes
to lower flux for such bursts, the brighter bursts redshifted out of the
$E_{peak}$ range being examined, and no bursts enter this band
through a redshift from a higher value of $E_{peak}$.  As a
consequence, the peak-flux distribution deviates from the $-5/2$ power law
in a manner determined by the luminosity distribution function for
$E_p = E_{peak}$, and a luminosity distribution function
for this value of $E_p$ can be derived.  One can then subtract from
the density of bursts with smaller values of $E_{peak}$ the density of
bursts with large values of $E_p$ that are strongly
redshifted.  One then selects bursts with a lower value of $E_{peak}$ to
repeat the process.  The uncertainty in the feasibility of this
analysis is in whether the BATSE data set is of sufficient size and
quality to perform such a test.


\section{Summary of Major Points}

A monoluminous burst distribution produces good fits to the
BATSE peak-flux distribution for very specific spectral parameters.
The parameters that produce the best values of $\chi^2$ are themselves
strongly dependent on the value of~$q_0$.  The spectra that are nearly
power laws do not produce good fits: a curved spectrum is required to
reproduce the observations.  The spectral parameters derived from fits
to gamma-ray burst spectra produce peak-flux distributions that fit
the observed distribution well. From model fits to the peak-flux
distribution, one finds that gamma-ray bursts at $z = 1$ have peak
fluxes of $\approx 1$--$2 \, \cm^{-2} \, \s^{-1}$ for $q_0 = 0.5$ and
$\approx 2.5$--$4 \, \cm^{-2} \, \s^{-1}$ for
$q_0 = 0.1$.  I confirm the conclusion that the $q_0 = 0.5$ cosmology
cannot simultaneously reproduce the observed flux distribution and the
reported time dilation--burst flux correlations
(\markcite{Fenimore2}Fenimore \& Bloom 1995),
and I find that the $q_0 \ll 0.5$ cosmology can never reproduce the
observed time dilation--burst flux correlation, independent of the
fit to the peak flux distribution.

When gamma-ray bursts have a distribution of luminosities, the shape
of the low end of the peak-flux distribution can be determined by
the shape of the luminosity distribution
(\markcite{Meszaros}M\'esz\'aros \& M\'esz\'aros 1995).
If the luminosity distribution below some value is proportional to
$L^{-\beta}$, then the asymptotic flux distribution as $F \rightarrow 0$
is $\propto F^{-\beta}$ when $\beta \gtrsim 0.6$ for $q_0 = 0.5$ and
$\beta \gtrsim 1.15$ for $q_0 = 0.1$ when there is only mild source
density evolution.  The precise values on the right
side of these inequalities are dependent on the burst spectrum and the
density evolution of burst sources.
When the inequalities are violated,
the peak-flux distribution is that found for monoluminous models.
Equations (12) and (16) give these limits and their dependence on
source density evolution.  The
luminosity distribution can formally diverge as the luminosity goes to 0;
below a critical luminosity, which is set by the peak-flux threshold
of the observing instrument, the luminosity distribution no longer
contributes to the peak-flux distribution.  A power-law luminosity
distribution $\propto L^{-\beta}$ with an upper cutoff at $L_0$ produces
peak-flux distributions that fit the BATSE 3B data well for most spectral
parameters.  One finds that $\beta \approx 1.5$. This shows that the
luminosity distribution need not be a standard candle or a steep power
law to produce a good fit to the data.

When the power-law luminosity distribution defines the shape of the flux
distribution, so that $\beta \approx 1.5$, the gamma-ray bursts at a given
peak flux have a distribution of redshifts that is strongly peaked at
$z < 1$. As $\beta$ becomes smaller, the peak in the $z$ distribution
goes to infinity, and the distribution rises monotonically.
The distribution goes to zero at some maximum value for $z$ that is
set by the peak flux.  When $\beta \approx 1.5$, the average redshift
is a weak function of peak flux when the peak flux is small.
At the peak-flux threshold of BATSE, the maximum redshift that a burst
can have is $\gtrsim 3$, while the average redshift at threshold is $< 1$.
From the model fits to the peak flux distribution, the distribution
in $z$ has an upper limit of $z = 1$ when
$F_0 \gtrsim 5 \, \cm^{-2} \, \s^{-1}$, although this limit on $F_0$ is
poorly constrained. This implies that the most luminous gamma-ray bursts
have an isotropic luminosity of $\gtrsim 2 \times 10^{51} \, \ergs \, \s^{-1}$
in the $50 \, \keV$ to $300 \, \keV$ energy band for
$H_0 = 75 \, \km \, \s^{-1} \, \Mpc^{-1}$.  For a burst duration of
$10 \, \s$, the fluence is $\gtrsim 2 \times 10^{52} \, \ergs \, \s^{-1}$,
making solar mass objects implausible sources of gamma-ray bursts.

A distinction in the literature is often made between narrow and broad
luminosity distributions.  By narrow, many authors mean either
that the bursts are nearly monoluminous, which is a statement about
the burst physics, or that a high percentage of observed bursts have nearly
the same flux, which is a statement about the detector threshold.
A more useful distinction is between luminosity-dominated flux
distributions and density-dominated flux distributions.  In the former,
the shape of the flux distribution is determined by the shape of the
luminosity distribution, and the average redshift, which is weakly dependent
on peak flux, is $\approx 1$.  In the latter,
the shape of the peak-flux distribution is determined by the spatial
distribution of gamma-ray bursts, and there is nearly a one-to-one
correspondence between redshift and peak flux.

The large errors in the model parameters do not strongly constrain the
distribution in $z$ at a given peak flux.  Because of this, large values of
$z$ found from model fits of the Compton attenuation spectrum to BATSE
gamma-ray bursts are consistent with the values of $z$ possible in the
flux distribution produced by a cosmological model with a luminosity
distribution function.

Because one averages over the distribution in $z$ when calculating the
average burst duration and the characteristic photon energy at a given
peak flux, the average time dilation factor does not equal the average
redshift factor; the former is larger than the latter.  From the model
fits to the observed peak-flux distribution, one expects bursts with two
different values of peak flux, one 10 times the other, to have ratios
of average time dilations and ratios of average spectral redshifts of
$< 1.6$.  Independent of the best fit to the flux distribution, the
maximum possible time dilation and redshift ratios are $< 1.7$.  The
reported correlations of $E_{peak}$ and burst time scale with peak flux
are much larger than this by 1.5 to 2 standard deviations, so they are
not consequences of cosmological expansion in the models examined above.
Within this model, only the addition of an intrinsic correlation with
luminosity can create the observed correlations.  An alternative not
explored in this article is strong source evolution, with
$\alpha \approx 3$ (\markcite{Reichart}Reichart \& M\'esz\'aros 1997).
From equations $(12)$ and $(16)$, the flux distribution in this case
is density-dominated.  On the other hand, since cosmological expansion
produces a modest effect when a luminosity distribution determines the
peak flux distribution, the absence of a correlation cannot be take as
evidence that gamma-ray bursts are not cosmological.

The presence of an intrinsic correlation of $E_{peak}$ with flux
may be tested by examining the flux distribution for bursts with
different values of $E_{peak}$, the energy at which the $\nu F_{\nu}$
curve has a maximum.  If the intrinsic value of $E_{peak}$
decreases with luminosity, then the bursts with the smallest observed
values of $E_{peak}$ will have a peak-flux distribution that falls away
from a $-5/2$ power law at a lower value of $F$ than do the bursts with
a high value of $E_{peak}$.  This is observed when gamma-ray
bursts are selected by hardness ratio, which is a proxy
for~$E_{peak}$.

One justification given for more sensitive gamma-ray burst detectors
is the measurement of the gamma-ray burst flux distribution at fluxes
below the thresholds of current gamma-ray experiments.  If gamma-ray bursts
are at cosmological redshifts of $z \approx 1$, then such an effort is of
limited value, because the behavior of the distribution is strongly model
dependent, and is easily modeled by a luminosity distribution function.
Of more value are large numbers of gamma-ray bursts with well characterized
spectra.  Such burst catalogs allow one to discern the dependence of burst
luminosity on spectral parameters, which then allows one to disentangle
the intrinsic correlation of burst luminosity with $E_{peak}$
from the correlation from the cosmological expansion.  It is this
that justifies accumulating as large a data set as possible, and underlies
the value of continued observations by the BATSE.


\end{document}